\documentclass[11pt,a4paper,epsf,epsfig,psfrag]{article}
\usepackage{jheppub}
\usepackage{amsmath,graphicx,amsfonts, mathrsfs,amssymb}
\usepackage{amsmath,epsfig}
\usepackage{amssymb,amsfonts}
\usepackage{color}
\usepackage{latexsym}
\usepackage{epsfig}
\usepackage{pdfsync}
\usepackage{marvosym}
\usepackage[utf8]{inputenc}
\newbox\pippobox

\title{Flavor dependent Critical endpoint from holographic QCD through machine learning}
\author[a, b]{Xun Chen}
\author[c]{and Mei Huang}

\affiliation[a]{School of Nuclear Science and Technology, University of South China, Hengyang 421001, China}
\affiliation[b]{Key Laboratory of Quark and Lepton Physics (MOE), Central China Normal University, Wuhan 430079,China.}
\affiliation[c]{School of Nuclear Science and Technology, University of Chinese Academy of Sciences, Beijing 100049, China}
\emailAdd{chenxun@usc.edu.cn}
\emailAdd{huangmei@ucas.ac.cn}

\date{\today}

\abstract{QCD phase diagram in the $T - \mu$ plane and the equation of state for pure gluon, 2-flavor, 2+1-flavor systems, and 2+1+1-flavor systems have been investigated using the Einstein-Maxwell-Dilaton (EMD) framework at finite temperature and chemical potential. By inputting lattice QCD data for the equation of state and baryon number susceptibility at zero chemical potential into holographic model, all the parameters can be determined with the aid of machine learning algorithms. Our findings indicate that the deconfinement phase transition is of first order for the pure gluon system with critical temperature $T_c = 0.265$ GeV at vanishing chemical potential. The phase transition for the 2-flavor, 2+1-flavor systems, and 2+1+1-flavor systems are crossover at vanishing chemical potential and first-order at high  chemical potential, and the critical endpoint (CEP) in the $T - \mu$ plane locates at ($\mu_B^c$=0.46 GeV, $T^c$=0.147 GeV), ($\mu_B^c$ = 0.74 GeV, $T^c$ = 0.094 GeV), and ($\mu_B^c$= 0.87 GeV,$T^c$ = 0.108 GeV), respectively. Additionally, the thermodynamic quantities of the system for different flavors at finite chemical potential are presented in this paper. It is observed that the difference between the 2+1-flavor and 2+1+1-flavor systems is invisible at vanishing chemical potential and low temperature. The location of CEP for 2+1+1-flavor system deviates explicitly from that of the 2+1-flavor system with the increase of chemical potential. Both 2+1-flavor and 2+1+1-flavor systems differ significantly from the 2-flavor system. Moreover, at zero temperature, the critical chemical potential is found to be $\mu_B$ = 1.1 GeV, 1.6 GeV, 1.9 GeV for the 2-flavor, 2+1-flavor and 2+1+1-flavor systems, respectively.}

\begin{document}
\maketitle

\section{Introduction}\label{sec:01_intro}
The exploration of the high-energy domain has been propelled forward by the Large Hadron Collider (LHC) at the European Organization for Nuclear Research (CERN), which has been in operation since 2009 \cite{ALICE:2010suc,ATLAS:2011ah,CMS:2012xss}. It has extended our experimental knowledge of the QCD phase diagram in the direction of high temperature, getting closer to the beginning of the universe. The next frontier on the phase diagram is the high-density regime \cite{Fukushima:2010bq}, where first principles calculations are known to encounter the fermion sign problem \cite{deForcrand:2009zkb}. Estimations based on the chiral model suggest that the quark-hadron transition turns from a crossover to a first-order phase transition at a finite baryon chemical potential, implying the existence of a critical point \cite{Pisarski:1983ms,Asakawa:1989bq,Stephanov:1998dy,Hatta:2002sj,Stephanov:1999zu,Hatta:2003wn,Schwarz:1999dj,Zhuang:2000ub}.

The equation of state (EoS) constitutes a cornerstone relation in the realm of thermodynamic quantities. The inception of explorations into the QCD EoS can be traced back to the pioneering MIT bag model \cite{Chodos:1974je}. Subsequently, an enriched comprehension of QCD thermodynamics has unfolded through the introduction of various model-based approaches, such as the potential model \cite{DeRujula:1975qlm} and the Nambu-Jona-Lasinio (NJL) model \cite{Nambu:1961tp,Nambu:1961fr} and their extensions. Theoretical endeavors have surged in recent years to probe the QCD phase transition and the associated EoS under  conditions of finite temperature and chemical potential. Noteworthy among these are the Dyson-Schwinger equations (DSE) \cite{Gao:2016qkh,Qin:2010nq,Shi:2014zpa,Fischer:2014ata}, the Polyakov-Nambu-Jona-Lasinio (PNJL) model \cite{McLerran:2008ua,Sasaki:2010jz,Li:2018ygx,Sun:2023kuu,Bao:2024glw}, the functional renormalization group (FRG) \cite{Fu:2019hdw,Zhang:2017icm}, and the hadron resonance gas models \cite{Becattini:2016xct,Fujimoto:2021xix}, among others. These methodologies collectively contribute to a progressive elucidation of the nuanced QCD landscape.

The holographic gauge-gravity duality has been motivated by the framework of string theory, which originally had an old and deep relationship with the strong interaction. Since its original proposal by Maldacena in 1997 \cite{Maldacena:1997re}, the holographic gauge-gravity duality has established itself as one of the breakthroughs in theoretical physics in the last few decades, being applied to obtain several insights into the non-perturbative physics of different strongly coupled quantum systems, including studies in the context of the strong interaction \cite{Erdmenger:2007cm,Brodsky:2014yha,Casalderrey-Solana:2011dxg,Adams:2012th}. Holographic gauge-gravity models are generally classified as either top-down constructions or bottom-up constructions. These models often incorporate phenomenological inputs and considerations to obtain a closer description of various aspects of real-world physical systems \cite{Rougemont:2023gfz}.

Except for the V-QCD model \cite{Jarvinen:2022doa}, the EMD model is a widely used bottom-up approach \cite{Gubser:2008ny,DeWolfe:2010he,He:2013qq,Yang:2014bqa,Yang:2015aia,Dudal:2017max,Dudal:2018ztm,Fang:2015ytf,Li:2022erd,Critelli:2017oub,Grefa:2021qvt,Arefeva:2020vae,Chen:2018vty,Chen:2020ath,Zhou:2020ssi,Chen:2019rez,Wang:2024szr}. This approach can quantitatively describe the strongly coupled Quark-Gluon Plasma (QGP) under extreme conditions. There are two predominant methods to obtain solutions of the EMD model. The first involves inserting the form of the dilaton potential and numerically constructing a family of five-dimensional black holes \cite{Gubser:2008ny,DeWolfe:2010he,Critelli:2017oub,Grefa:2021qvt,He:2022amv,Fu:2024wkn,Liu:2023pbt,Liu:2024efy,Cao:2024jgt}. The second method, known as the potential reconstruction method \cite{He:2013qq,Yang:2014bqa,Yang:2015aia,Dudal:2017max,Fang:2015ytf,Chen:2018vty,Chen:2020ath,Zhou:2020ssi,Jokela:2024xgz}, involves inputting the dilaton or a metric function to determine the dilaton potential. Although this approach results in a temperature-dependent dilaton potential, the model can nevertheless capture many QCD properties through analytical solutions.

\begin{figure}
    \centering
    \includegraphics[width=12cm]{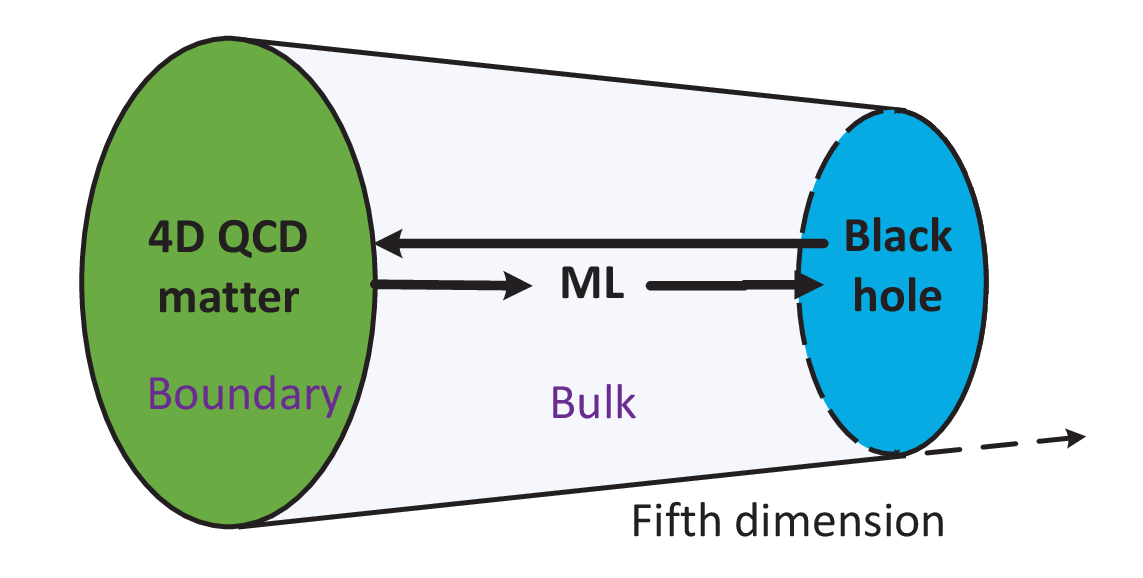}
    \caption{\label{sketch0} The sketch of holographic QCD and machine learning.}
\end{figure}

Machine learning has become a useful tool in high-energy physics; for a recent review, see Ref. \cite{Zhou:2023pti}. Furthermore, the integration of deep learning with holographic QCD has been explored in recent studies \cite{Hashimoto:2018ftp,Akutagawa:2020yeo,Hashimoto:2018bnb,Yan:2020wcd,Hashimoto:2021ihd,Song:2020agw,Chang:2024ksq,Ahn:2024gjf,Gu:2024lrz,Li:2022zjc}. Unlike conventional holographic models, this approach first employs specific QCD data to determine the bulk metric (as well as other model parameters) through machine learning. Subsequently, the model utilizes the determined metric to calculate other physical QCD observables, serving as predictions of the model. A sketch of the holographic QCD and machine learning is shown in Fig. \ref{sketch0}. Following our previous letter \cite{Chen:2024ckb}, this paper will give a detailed study of QCD phase diagram with input of lattice information with the aid of machine learning at finite temperature and chemical potential. The parameter space is high-dimensional and non-linear, making traditional fitting methods less effective. Our work wants to develop an effectively automatic way through machine learning to determine the parameters in the theoretical model.

In this study, we will employ the potential reconstruction method, augmented by machine learning, to develop an EMD model. This model will then be used to predict the CEP. The potential reconstruction method in principle should be equivalent to the framework of DeWolfe-Gubser-Rosen (DGR) \cite{Gubser:2008ny,DeWolfe:2010he}. In the potential reconstruction framework, the metric $A(z)$ can be obtained through machine leaning the lattice EOS, then the dilaton potential $V(\Phi)$ can be solved, and in the DGR framework, the dilaton potential $V(\Phi)$ can be obtained through the lattice EOS, and then $A(z)$ can be solved. A significant advantage of the potential reconstruction method in our model is the ability to derive analytical solutions. Besides, compared with the V-QCD model and DGR model, we have fewer parameters in our model by machine learning.

The remainder of the paper is organized as follows: In Sec.~\ref{sec:02_setup}, we review the holographic EMD model. We show detailed process of parameter determination using machine learning techniques in Sec.~\ref{sec:03} . In Sec.~\ref{sec:04}, we compare the EoS of our model with lattice for different flavors at vanishing chemical potential. Sec.~\ref{sec:05} is devoted to the calculation of the EoS at finite chemical potential for different flavors. The high-order baryon number susceptibilities are discussed in Sec.~\ref{sec:06}. The QCD phase diagram and CEP are discussed in Sec.~\ref{sec:07}. Finally, the conclusion and outlook are offered in Sec.~\ref{sec:08}.

\section{The review of EMD model}\label{sec:02_setup}

Firstly, we review the 5-dimensional Einstein-Maxwell-Dilaton system \cite{He:2013qq,Yang:2014bqa,Yang:2015aia,Dudal:2017max,Dudal:2018ztm,Chen:2018vty,Chen:2020ath,Zhou:2020ssi,Chen:2019rez}. The action includes a gravity field $g_{\mu \nu}^s$, a Maxwell field $A_\mu$ (we set $A_i=0$ and make a gauge choice $A_z=0$) and a neutral dilatonic scalar field. In the string frame, it is expressed by the following equation
\begin{equation}
S_b=\frac{1}{16 \pi G_5} \int d^5 x \sqrt{-g^s} e^{-2 \phi_s}\left[R_s-\frac{f_s\left(\phi_s\right)}{4} F^2+4 \partial_\mu \phi_s \partial^\mu \phi_s-V_s\left(\phi_s\right)\right],
\label{Eq:actionsb}
\end{equation}
where $f(\phi)$ is the gauge kinetic function coupled with the Maxwell field $A_\mu$, $V\left(\phi\right)$ is the potential of the dilaton field and $G_5$ is the Newton constant in five dimensions. The explicit forms of the gauge kinetic function $f\left(\phi\right)$ and the dilaton potential $V\left(\phi\right)$ can be solved consistently from the equations of motion (EoMs).
To study the thermodynamics of QCD, we transform the action from string frame to Einstein frame with the following transformations,
\begin{equation}
\phi_s=\sqrt{\frac{3}{8}} \phi, \quad g_{\mu \nu}^s=g_{\mu \nu} e^{\sqrt{\frac{2}{3}} \phi}, \quad f_s\left(\phi_s\right)=f(\phi) e^{\sqrt{\frac{2}{3}} \phi}, \quad V_s\left(\phi_s\right)=e^{-\sqrt{\frac{2}{3}} \phi} V(\phi).
\end{equation}
The action in Einstein frame becomes
\begin{equation}
\begin{aligned}
S_b & =\frac{1}{16 \pi G_5} \int d^5 x \sqrt{-g}\left[R-\frac{f(\phi)}{4} F^2-\frac{1}{2} \partial_\mu \phi \partial^\mu \phi-V(\phi)\right]. \\
\end{aligned}
\end{equation}
EoMs derived from the action are, respectively,
\begin{equation}
\begin{gathered}
R_{M N}-\frac{1}{2} g_{M N} R-T_{M N}=0, \\
\nabla_M\left[f(\phi) F^{M N}\right]=0, \\
\partial_M\left[\sqrt{-g} \partial^M \phi\right]-\sqrt{-g}\left(\frac{\partial V}{\partial \phi}+\frac{F^2}{4} \frac{\partial f}{\partial \phi}\right)=0,
\end{gathered}
\end{equation}
where
\begin{equation}
\begin{aligned}
T_{M N}= & \frac{1}{2}\left(\partial_M \phi \partial_M \phi-\frac{1}{2} g_{M N}(\partial \phi)^2-g_{M N} V(\phi)\right) +\frac{f(\phi)}{2}\left(F_{M P} F_N^P-\frac{1}{4} g_{M N} F^2\right) .
\end{aligned}
\end{equation}
We give the following ansatz of metric
\begin{equation}
d s^2=\frac{L^2 e^{2 A(z)}}{z^2}\left[-g(z) d t^2+\frac{d z^2}{g(z)}+d \vec{x}^2\right],
\end{equation}
where $z$ is the 5th-dimensional holographic coordinate and the radius $L$ of $\rm AdS_5$ space is set to be one. Using the above ansatz of the metric, the EoMs and constraints for the background fields can be obtained as
\begin{equation}
\begin{gathered}
\phi^{\prime \prime}+\phi^{\prime}\left(-\frac{3}{z}+\frac{g^{\prime}}{g}+3 A^{\prime}\right)-\frac{L^2 e^{2 A}}{z^2 g} \frac{\partial V}{\partial \phi}+\frac{z^2 e^{-2 A} A_t^{\prime 2}}{2 L^2 g} \frac{\partial f}{\partial \phi}=0,
\end{gathered}
\end{equation}
\begin{equation}
A_t^{\prime \prime}+A_t^{\prime}\left(-\frac{1}{z}+\frac{f^{\prime}}{f}+A^{\prime}\right)=0,
\end{equation}
\begin{equation}
g^{\prime \prime}+g^{\prime}\left(-\frac{3}{z}+3 A^{\prime}\right)-\frac{e^{-2 A} A_t^{\prime 2} z^2 f}{L^2}=0,
\end{equation}
\begin{equation}
\begin{aligned}
A^{\prime \prime} & +\frac{g^{\prime \prime}}{6 g}+A^{\prime}\left(-\frac{6}{z}+\frac{3 g^{\prime}}{2 g}\right)-\frac{1}{z}\left(-\frac{4}{z}+\frac{3 g^{\prime}}{2 g}\right)+3 A^{\prime 2} +\frac{L^2 e^{2 A} V}{3 z^2 g}=0,
\end{aligned}
\end{equation}
\begin{equation}
A^{\prime \prime}-A^{\prime}\left(-\frac{2}{z}+A^{\prime}\right)+\frac{\phi^{\prime 2}}{6}=0,
\end{equation}
where only four of the above five equations are independent.
The boundary conditions near the horizon are
\begin{equation}
A_t\left(z_h\right)=g\left(z_h\right)=0.
\end{equation}
Near the boundary $z \rightarrow z_h$, we require the metric in the string frame to be asymptotic to $\rm AdS_5$. The boundary conditions are
\begin{equation}
A(0)=-\sqrt{\frac{1}{6}} \phi(0), \quad g(0)=1, \quad A_t(0)=\mu+\rho^{\prime} z^2+\cdots.
\end{equation}
$\mu$ can be regarded as baryon chemical potential and $\rho^{\prime}$ is proportional to the baryon number density. $\mu$ is related to the quark-number chemical potential $\mu = 3\mu_q$. The baryon number density can be calculated as \cite{Critelli:2017oub,Zhang:2022uin}
\begin{equation}
\begin{aligned}
\rho & =\left|\lim _{z \rightarrow 0} \frac{\partial \mathcal{L}}{\partial\left(\partial_z A_t\right)}\right| \\
& =-\frac{1}{16\pi G_5} \lim _{z \rightarrow 0}\left[\frac{\mathrm{e}^{A(z)}}{z} f(\phi) \frac{\mathrm{d}}{\mathrm{d} z} A_t(z)\right].
\end{aligned}
\end{equation}
$\mathcal{L}$ is the Lagrangian density in the Einstein frame. The EoMs can be analytically solved as
\begin{equation}
\begin{aligned}
\phi^{\prime}(z) & =\sqrt{-6\left(A^{\prime \prime}-A^{\prime 2}+\frac{2}{z} A^{\prime}\right)}, \\
A_t(z) & =\sqrt{\frac{-1}{\int_0^{z_H} y^3 e^{-3 A} d y \int_{y_g}^y \frac{x}{e^A f} d x}} \int_{z_h}^z \frac{y}{e^A f} d y, \\
g(z) & =1-\frac{\int_0^z y^3 e^{-3 A} d y \int_{y_g}^y \frac{x}{e^A f} d x}{\int_0^{z_h} y^3 e^{-3 A} d y \int_{y_g}^y \frac{x}{e^A f} d x}, \\
V(z) & =-3 z^2 g e^{-2 A}\left[A^{\prime \prime}+3 A^{\prime 2}+\left(\frac{3 g^{\prime}}{2 g}-\frac{6}{z}\right) A^{\prime}-\frac{1}{z}\left(\frac{3 g^{\prime}}{2 g}-\frac{4}{z}\right)+\frac{g^{\prime \prime}}{6 g}\right].
\end{aligned}
\end{equation}
The only undetermined integration constant $y_g$ can be related to the chemical potential $\mu$ in the following way. We expand the field $A_t(z)$ near the boundary at $z=0$ to obtain
\begin{equation}
A_t(0)=\sqrt{\frac{-1}{\int_0^{z_h} y^3 e^{-3 A} d y \int_{y_g}^y \frac{x}{e^A f} d x}}\left(-\int_0^{z_h} \frac{y}{e^A f} d y+\frac{1}{e^{A(0)} f(0)} z^2+\cdots\right) .
\end{equation}
From the AdS/CFT dictionary, we can define the chemical potential in the system as
\begin{equation}
\mu=-\sqrt{\frac{-1}{\int_0^{z_H} y^3 e^{-3 A} d y \int_{y_g}^y \frac{x}{e^A f} d x}} \int_0^{z_H} \frac{y}{e^A f} d y.
\end{equation}
The gauge kinetic function $f(z)$ is taken as
\begin{equation}\label{fff}
f(z)=e^{c z^2-A(z)+k}.
\end{equation}
The integration constant $y_g$ in term of the chemical potential $\mu$ is
\begin{equation}
e^{c y_g^2}=\frac{\int_0^{z_H} y^3 e^{-3 A-c y^2} d y}{\int_0^{z_H} y^3 e^{-3 A} d y}+\frac{\left(1-e^{c z_h^2}\right)^2}{2 c \mu^2 e^k \int_0^{z_h} y^3 e^{-3 A} d y}.
\end{equation}
Then, we can get
\begin{equation}
\begin{aligned}
g(z)&=1-\frac{1}{\int_0^{z h} d x x^3 e^{-3 A(x)}}\left[\int_0^z d x x^3 e^{-3 A(x)}+\frac{2 c \mu^2 e^k}{\left(1-e^{-c z_h^2}\right)^2} \operatorname{det} \mathcal{G}\right],\\
\phi^{\prime}(z) & =\sqrt{6\left(A^{\prime 2}-A^{\prime \prime}-2 A^{\prime} / z\right)}, \\
A_t(z) & =\mu \frac{e^{-c z^2}-e^{-c z_h^2}}{1-e^{-c z_h^2}}, \\
V(z) & =-\frac{3 z^2 g e^{-2 A}}{L^2}\left[A^{\prime \prime}+A^{\prime}\left(3 A^{\prime}-\frac{6}{z}+\frac{3 g^{\prime}}{2 g}\right)-\frac{1}{z}\left(-\frac{4}{z}+\frac{3 g^{\prime}}{2 g}\right)+\frac{g^{\prime \prime}}{6 g}\right],
\end{aligned}
\end{equation}
where
\begin{equation}
\operatorname{det} \mathcal{G}=\left|\begin{array}{ll}
\int_0^{z_h} d y y^3 e^{-3 A(y)} & \int_0^{z_h} d y y^3 e^{-3 A(y)-c y^2} \\
\int_{z_h}^z d y y^3 e^{-3 A(y)} & \int_{z_h}^z d y y^3 e^{-3 A(y)-c y^2}
\end{array}\right|.
\end{equation}
The Hawking temperature and entropy of this black hole solution are given by,
\begin{equation}
\begin{aligned}
T & =\frac{z_h^3 e^{-3 A\left(z_h\right)}}{4 \pi \int_0^{z_h} d y y^3 e^{-3 A(y)}}\Big[1+ \\
&\frac{2 c \mu^2 e^k\left(e^{-c z_h^2} \int_0^{z_h} d y y^3 e^{-3 A(y)}-\int_0^{z_h} d y y^3 e^{-3 A(y)} e^{-c y^2}\right)}{(1-e^{-c z_h^2})^2} \Big],
\end{aligned}
\end{equation}
\begin{equation}\label{SSS}
S_{\mathrm{BH}}=\frac{e^{3 A\left(z_h\right)}}{4 G_5 z_h^3}.
\end{equation}
To obtain an analytical solution to the system, we assume
\begin{equation}\label{AAA}
A(z)= d*\ln(a z^2 + 1) + d*\ln(b z^4 + 1),
\end{equation}
and in the string frame it takes the form of
\begin{equation}
A_s(z)=A(z)+\sqrt{\frac{1}{6}} \phi(z).
\end{equation}
with five undetermined parameters, which will be fixed by the input of the EoS from lattice results with the help of machine learning. Moreover, there is a parameter $c$ which can be determined from the baryon number susceptibility at the same time.  After knowing the entropy, the free energy can be calculated as
\begin{equation}
\begin{aligned}
F&=-\int s d T-\int \rho d \mu=-\int s\left(\frac{\partial T}{\partial z_h} d z_h+\frac{\partial T}{\partial \mu} d \mu\right)-\int \rho d \mu\\
&= \int_{z_h}^\infty s\frac{\partial T}{\partial z_h} d z_h-\int_0^\mu (s\frac{\partial T}{\partial \mu} d \mu\ +\rho) d \mu.
\end{aligned}
\end{equation}
We have normalized the free energy to vanish at $z_h \rightarrow \infty$. The energy density of the system can be derived as
\begin{equation}
\epsilon=-p+s T+\mu \rho.
\end{equation}
The specific heat can be defined as
\begin{equation}
C_V=T \frac{\partial s}{\partial T}.
\end{equation}
For non-zero chemical potential conditions, the squared speed of sound can be calculated by \cite{Li:2020hau,Yang:2017oer,Gursoy:2017wzz}
\begin{equation}
C_s^2=\frac{s}{T\left(\frac{\partial s}{\partial T}\right)_\mu+\mu\left(\frac{\partial \rho}{\partial T}\right)_\mu}.
\end{equation}
We also numerically verified that this definition matches the one given by Eq. A10 in Appendix A of Ref. \cite{Floerchinger:2015efa} at constant particle density. The second-order baryon number susceptibility is defined as
\begin{equation}
\chi_2^B=\frac{1}{T^2} \frac{\partial \rho}{\partial \mu}.
\end{equation}
The higher-order baryon number susceptibility can be defined as \cite{Li:2023mpv}
\begin{equation}
\chi_n^B=\frac{\partial^n}{\partial (\mu_B/T)^n} \frac{P}{T^4}.
\end{equation}

\section{Determining the parameters with machine learning}\label{sec:03}
\begin{figure}
    \centering
    \includegraphics[width=16cm]{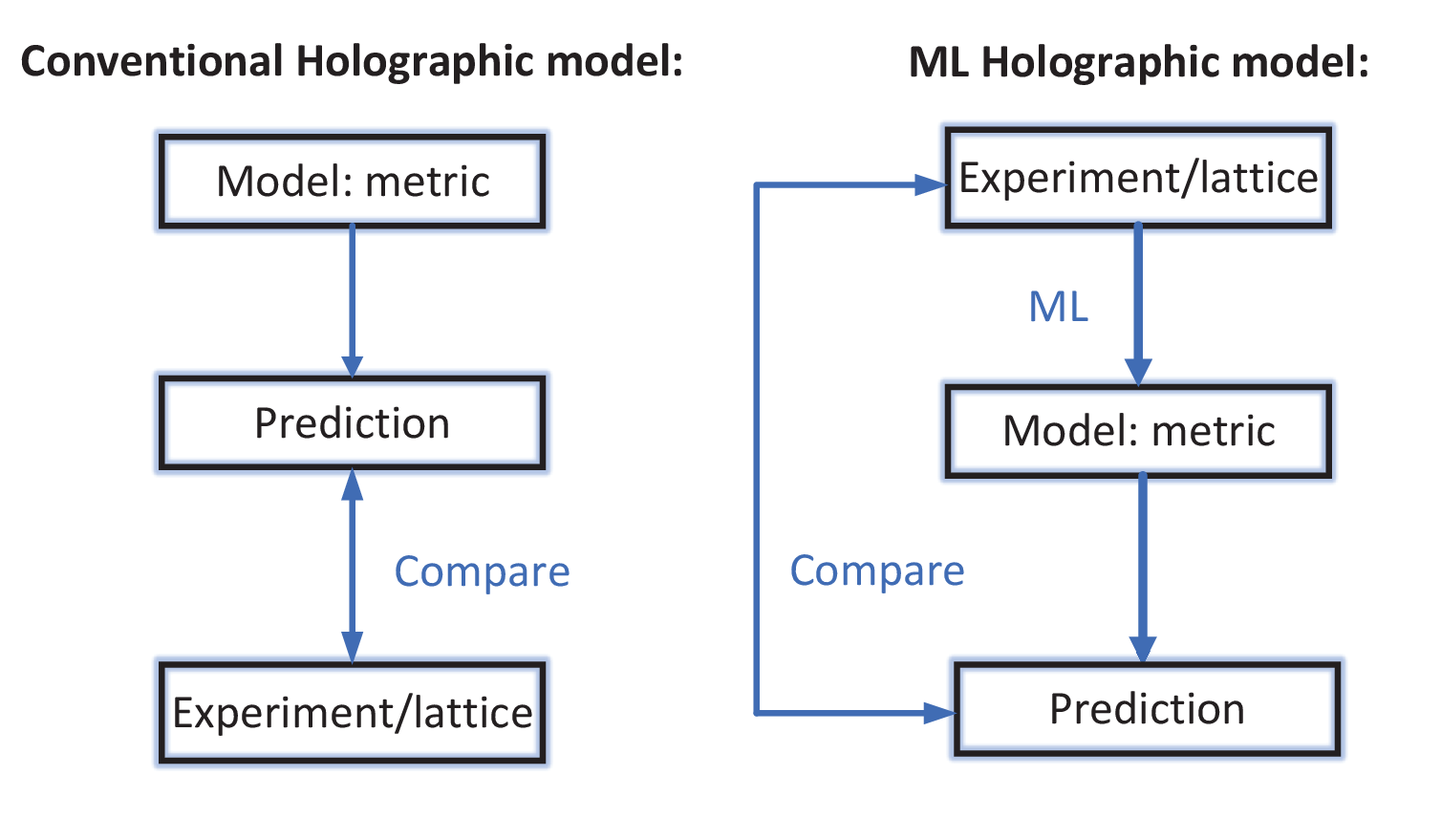}
    \caption{\label{sketch1} The difference between ML holographic model and the traditional holographic method.}
\end{figure}

In this section, we will describe how to determine all the parameters of the model using machine learning. We illustrate the key differences between our model and the traditional method visually in Fig. \ref{sketch1}. Our model comprises six parameters: $a$, $b$, $c$, $d$, $k$ and $G_5$. These parameters are associated with $A(z)= d*\ln(a z^2 + 1) + d*\ln(b z^4 + 1)$ in Eq.~(\ref{AAA}), $f(z)=e^{c z^2-A(z)+k}$ in Eq.~(\ref{fff}) and $S_{\mathrm{BH}}=\frac{e^{3 A\left(z_h\right)}}{4 G_5 z_h^3}$ in Eq.~(\ref{SSS}). We can simultaneously constrain all these parameters by inputting the EoS and baryon number susceptibility derived from lattice results. In the EoS, each thermodynamic quantity is interconnected, affecting the others. For our analysis, we specifically chose to represent entropy as a function of temperature. To illustrate the practical application of our model, we will present a case study using the 2+1-flavor scenario.

\begin{figure}
    \centering
    \includegraphics[width=16cm]{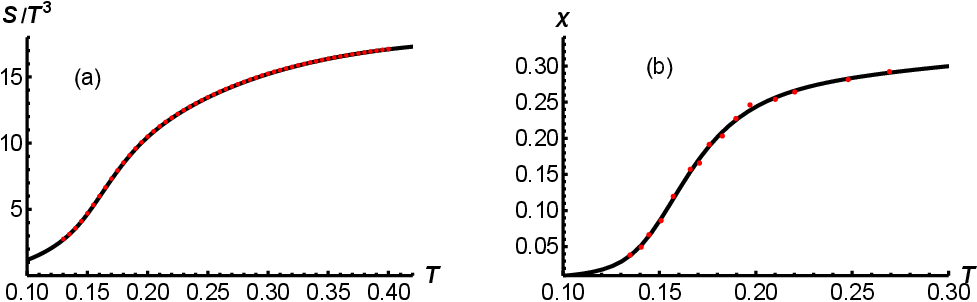}
    \caption{\label{learn}  (a) The entropy as a function of temperature. (b) The baryon number susceptibility as a function of temperature. The dots represent the results obtained from lattice simulations, while the black line indicates the prediction made by the neural network. The unit of $T$ is in $\rm GeV$.}
\end{figure}

First, we use "TensorFlow" to build a neural network model for regression tasks. It consists of several steps. We prepare a dataset comprising entropy and temperature data derived from lattice results, which serve as both input and target output variables. We employ a Sequential model, which consists of four fully connected layers. Each layer utilizes a sigmoid activation function. The first layer contains 64 nodes, all of which receive the input value $T$ through the sigmoid function. The second layer has 128 nodes, while the third layer returns to 64 nodes, all with sigmoid activation. The final layer is singular, consisting of one node that corresponds to the output variable, entropy. We define the loss function as mean squared error (MSE) to measure the model's prediction errors, and we select the Adam algorithm as our optimizer, setting the learning rate at 0.001 for efficient gradient descent. Following the training of the neural network model, we establish a predictive relationship between the input variable and the target output, enabling the model to forecast the target output with high accuracy on the test set. The trained model's performance is depicted in Fig. \ref{learn} (a), where it effectively captures the relationship between entropy and temperature, as evidenced by the lattice results from  \cite{HotQCD:2014kol}. Similarly, Fig. \ref{learn} (b) illustrates the model's success in representing the baryon number susceptibility as a function of $T$, with data sourced from Ref. \cite{Bazavov:2017dus}. These results confirm that our neural network model accurately describes the lattice data.

\begin{figure}
    \centering
    \includegraphics[width=16cm]{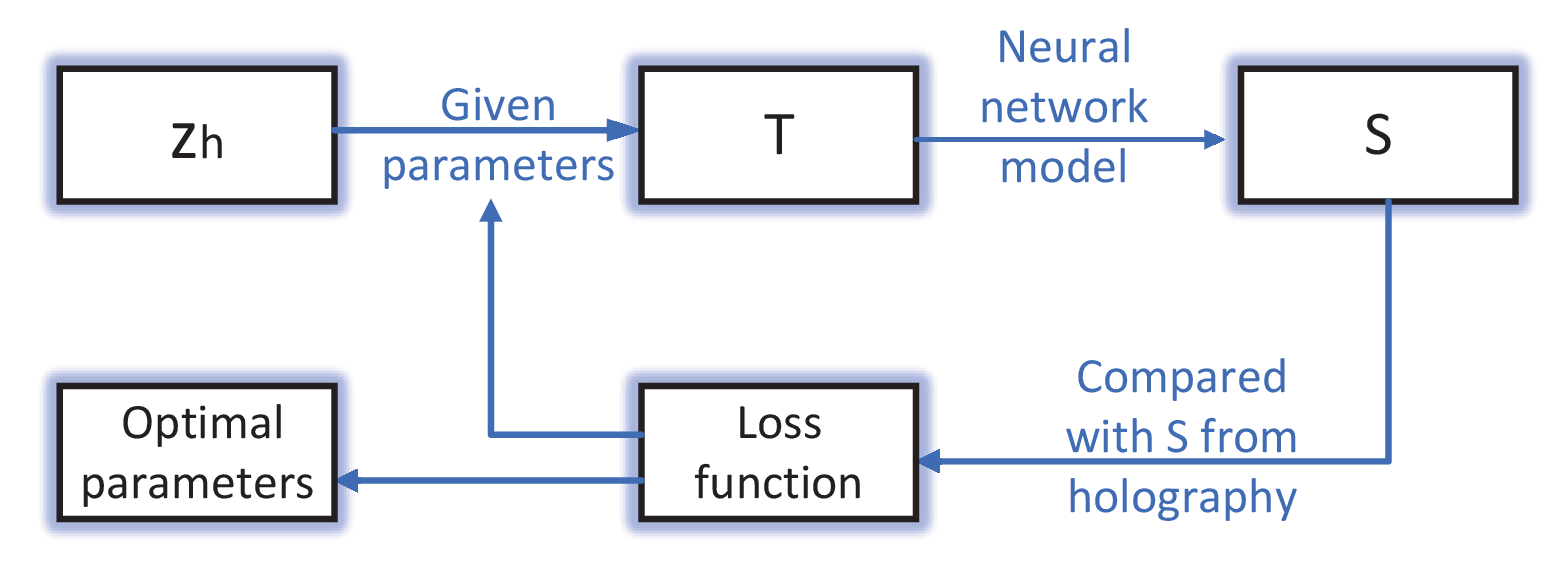}
    \caption{\label{sketch3} A sketch of the learning progress in our model.}
\end{figure}

Lastly, we transform the problem into an optimization task, leveraging a gradient descent algorithm to find the optimal parameter values. The functions $A(z)$, $T(z_h)$, $\chi(z_h)$ and $S(z_h)$  are defined on the holographic side. Then, we define the loss function, denoted as "$loss$($a$, $b$, $c$, $d$, $k$, $G_5$)", as the difference between the predictions of the holographic model and the values obtained from the neural network model, employing the least squares method to minimize this difference. Within the loss function, we compute $T_{test}$, $\chi_{pred}$ and $S_{pred}$ using the provided parameters ($a$, $b$, $c$, $d$, $k$, $G_5$), and then calculate the MSE between the predicted values and those derived from the neural network model. Initial parameter values should be defined, and the "Adam" optimizer is used for training. We apply constraints to our parameters, specifically  $a, b\geq0$ and $d\leq0$  to ensure that the solutions for $\phi(z)$ and $A(z)$ are real-valued. In every training epoch, we first calculate the loss function and then compute the gradients. With this information, we update the parameter values accordingly. Finally, we output the optimal parameter values obtained after training. Fig. \ref{sketch3} visually depicts this learning progress.

Note that the loss function in the final step comprises two components: the MSE between the predicted entropy and the actual entropy values, plus the MSE between the predicted baryon number susceptibility and the actual baryon number susceptibility values. We can adjust the weights to achieve a balanced fit between these two results. However, a perfect fit for both parts is not possible due to the choice of the coupling functions $f(\phi)$ and $A(z)$. Since this work aims to find an analytical solution, the selection of $A(z)$ and $f(\phi)$ can not be overly complex. Once all the parameters are determined, we can predict the critical temperature $T_c$ in Table \ref{table:parameter}. The QCD phase diagram in the $T-\mu$ plane will be explored in the following sections.

\begin{table}[htbp]
	\centering
	\begin{tabular}{|c|c|c|c|c|c|c|c|c|}
		\hline
		& $a$ & $b$ & $c$ & $d$ & $k$ & $G_5$ & $T_c$ (GeV) & CEP (GeV) \\
		\hline
		$N_f = 0$ & 0 & 0.072 & 0 & -0.584 & 0 & 1.326 & 0.265 & / \\
        \hline
		$N_f = 2$ & 0.067 & 0.023 & -0.377 & -0.382 & 0 & 0.885 & 0.189 & ($\mu_B^c$=0.46, $T^c$=0.147) \\
        \hline
        $N_f = 2+1$ & 0.204 & 0.013 & -0.264 & -0.173 & -0.824 & 0.400 & 0.128 & ($\mu_B^c$=0.74, $T^c$=0.094) \\
        \hline
        $N_f = 2+1+1$ & 0.196 & 0.014 & -0.362 & -0.171 & -0.735 & 0.391 & 0.131 & ($\mu_B^c$=0.87, $T^c$=0.108) \\
        \hline
	\end{tabular}
\caption{Parameters given by the machine learning of pure gluon system, 2 flavor, 2+1 flavor system, and 2+1+1 flavor system, respectively. $T_c$ is the critical temperature determined through the $c_s^2$ inflection for crossover at vanishing chemical potential. The parameters are from $A(z)= d*\ln(a z^2 + 1) + d*\ln(b z^4 + 1)$ and $f(z)=e^{c z^2-A(z)+k}$ defined in Eqs.~(\ref{AAA}) and~(\ref{fff}), and $G_5$ appears in the action Eq.~(\ref{Eq:actionsb}) which measures the entropy $S_{\mathrm{BH}}=\frac{e^{3 A\left(z_h\right)}}{4 G_5 z_h^3}$  Eq.~(\ref{SSS}). The unit of $T$ is GeV.}
\label{table:parameter}
\end{table}

\section{Thermodynamics at vanishing chemical potential for different flavors}\label{sec:04}

\begin{figure}
    \centering
    \includegraphics[width=16cm]{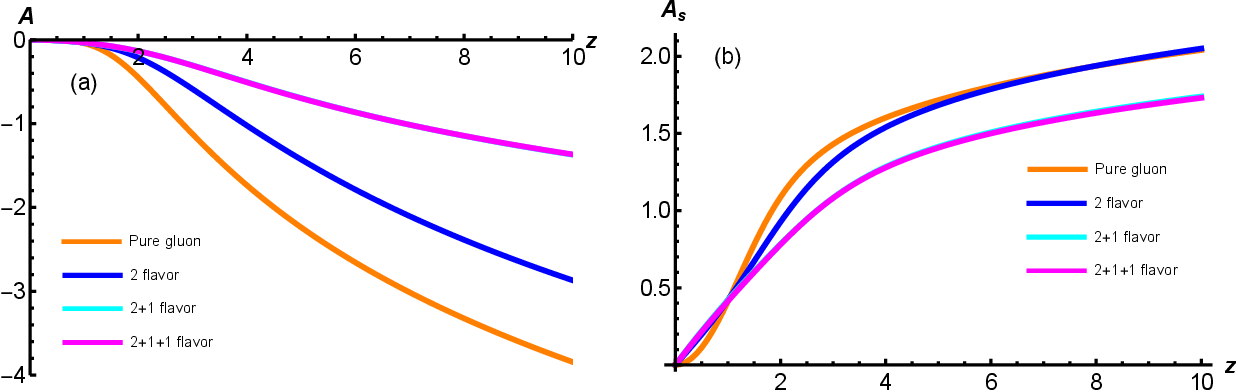}
    \caption{\label{comA}  (a) Comparison of $A$ for different flavors. (b) Comparison of $A_S$ for different flavors. The unit of $z$ is in $\rm GeV^{-1}$.}
\end{figure}

In this section, we compare different models for various systems. We start by examining the current model, whose parameters were detailed in the previous section. Using these parameters, we have depicted the deform factor of the metric $A$ and $A_S$ of our model in Fig.~\ref{comA}. It is observed that both $A$ and $A_S$ increase with the increasing number of flavors. However, the difference for both $A$ and $A_S$ between the 2+1-flavor and 2+1+1-flavor cases is almost invisible.

\begin{figure}
    \centering
    \includegraphics[width=16cm]{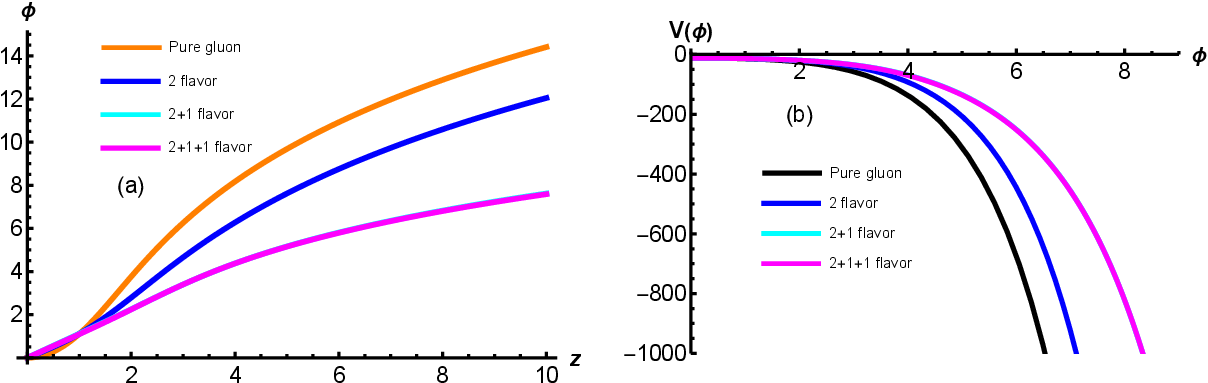}
    \caption{\label{comphi} (a) Comparison of the dilaton field $\phi$ for different flavors. (b) Comparison of the dilaton potential $V_\phi$ for different flavors. $T$ and $\mu$ are fixed as 0. The unit of $z$ is in $\rm GeV^{-1}$.}
\end{figure}

Fig.~\ref{comphi} presents the behavior of dilaton field $\phi$ and the dilaton potential $V_{\phi}$ for different flavors. It is shown that the absolute value of $V_{\phi}$ decrease with the increasing number of flavors. The difference between the 2+1-flavor and 2+1+1-flavor cases for $V_{\phi}$ is also invisible.

\begin{figure}
    \centering
    \includegraphics[width=16cm]{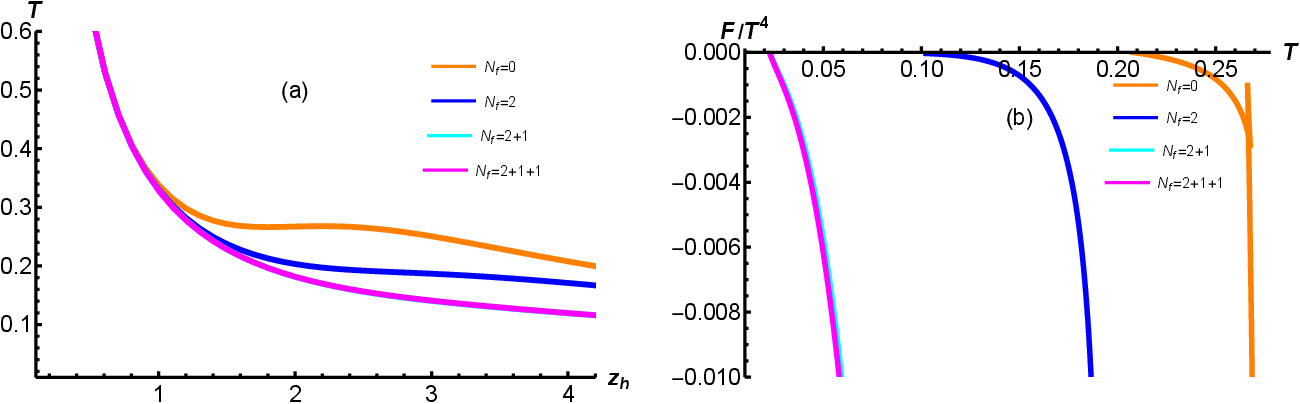}
    \caption{\label{mu0tem} (a) The temperature as a function of $z_h$ for different flavors. (b) The free energy as a function of temperature for different flavors. The unit of $z_h$ is $\rm GeV^{-1}$. The unit of temperature is $\rm GeV$. The unit of free energy is in $\rm GeV^4$.}
\end{figure}

Furthermore, we can depict the temperature and free energy for different flavors in our model in Fig.~\ref{mu0tem}. The temperature is only a multi-valued function of $z_h$ for pure gluon. For non-vanishing flavor, the temperature shows a smoothly decreasing function of $z_h$. The free energy, as deduced from entropy, manifests the expected swallowtail shape, indicating a critical temperature at $T_c = 0.265~\text{GeV}$, as shown in Fig.~\ref{mu0tem} (b). If we were to observe only the lattice QCD data, we could not conclude that the phase transition is first order. Our model provides additional predictions beyond the data obtained from lattice QCD. In other cases, there is no crossing point, and the phase transition is a crossover.

\begin{figure}
    \centering
    \includegraphics[width=16cm]{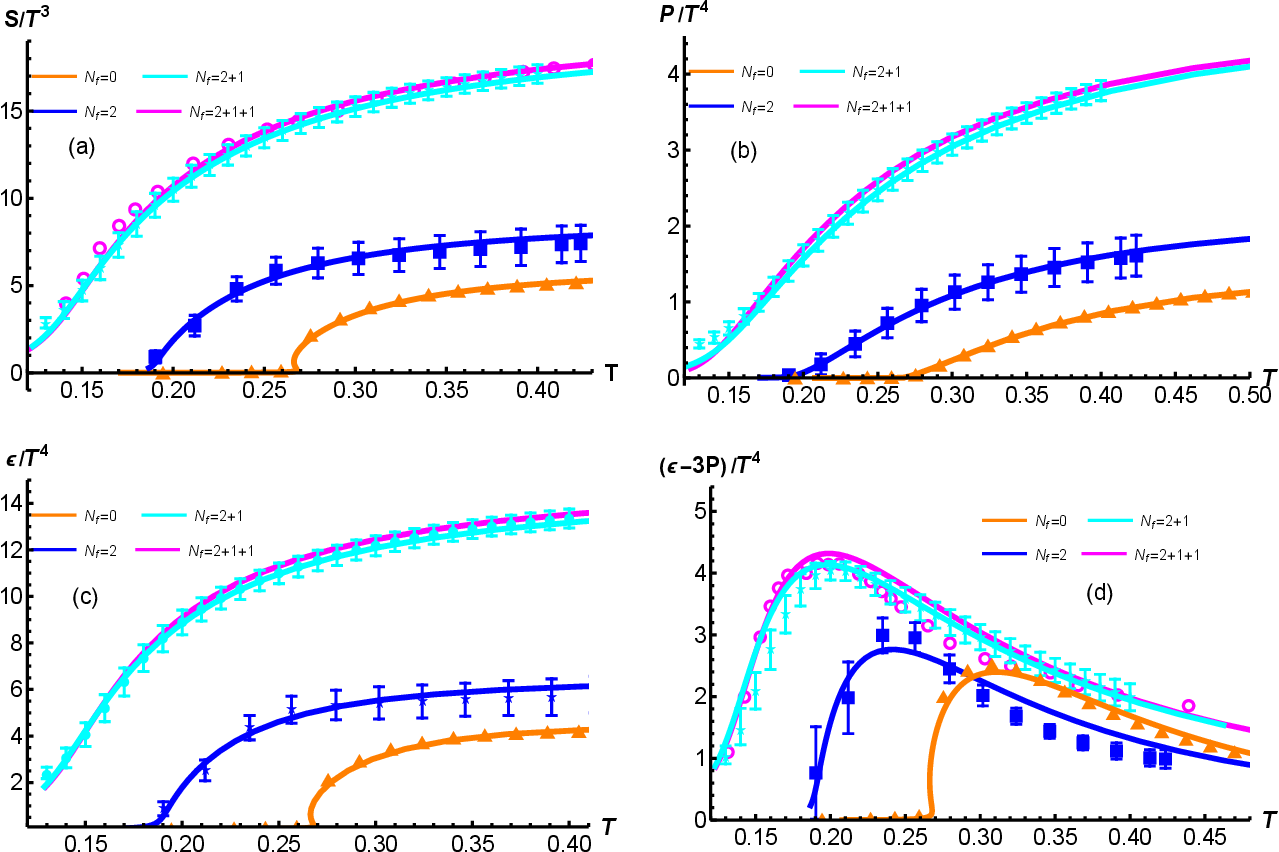}
    \caption{\label{mu0eos} (a) The entropy as a function of the temperature for different flavors in our model.  (b) The pressure as a function of temperature for different flavors in our model. (c) The energy as a function of temperature for different flavors in our model. (d) The trace anomaly as a function of temperature for the 2+1+1-flavor case in our model. Lattice results are taken from Refs. \cite{Borsanyi:2012ve,Burger:2014xga,HotQCD:2014kol,Ratti:2013uta}.}
\end{figure}

In Fig.~\ref{mu0eos}, we summarize the holographic EoS for all the flavors and compare them with lattice results. In our framework, the models closely fit the lattice results, indicating that our model can incorporate the EoS lattice data information for  $N_f = 0$, $N_f = 2$, $N_f = 2+1$, and $N_f= 2+1+1$ at a vanishing chemical potential. It is shown that the EoS for 2+1 flavor and 2+1+1 flavor is similar, since the effect of charm quark is not significant at low temperatures. At high temperatures, the difference between the 2+1 flavor and the 2+1+1 flavor becomes more pronounced.

\begin{figure}
    \centering
    \includegraphics[width=16cm]{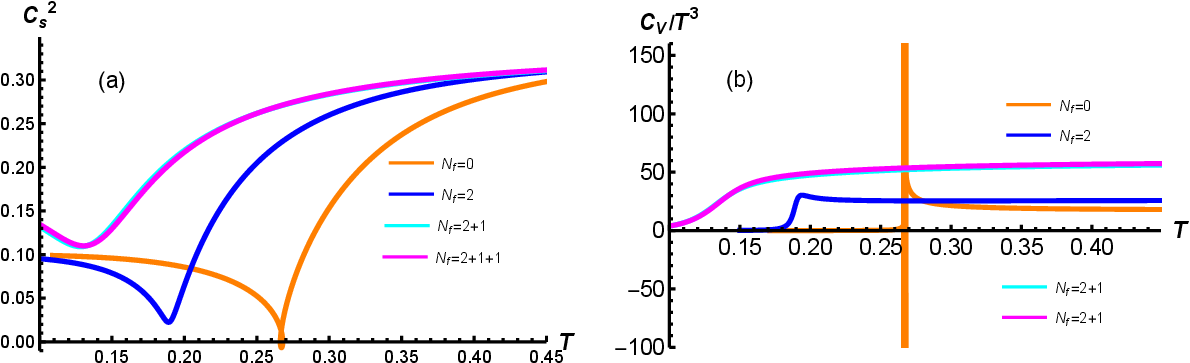}
    \caption{\label{mu0cscv} (a) The square of the speed of sound as a function of temperature. (b) The specific heat as a function of temperature. The unit of temperature is GeV. The unit of the specific heat is in $\rm GeV^3$.}
\end{figure}

Additionally, we calculate the square of the speed of sound, $C_s^2$, and the specific heat, $C_v$, as shown in Fig.~\ref{mu0cscv}. In the high-temperature limit, $C_s^2$ approaches the conformal value of $1/3$, as expected. The speed of sound becomes multi-valued around $T_c$ for the pure gluon. Furthermore, the emergence of a negative branch for the speed of sound leads to an imaginary value for the pure gluon, implying the Gregory-Laflamme dynamical instability \cite{He:2020fdi,Gregory:1993vy,Gregory:1994bj}. This dynamical instability is analogous to the thermodynamic instability inferred from the specific heat, as suggested by the Gubser-Mitra conjecture. The negative branch of the specific heat corresponds to the thermodynamic instability of the black hole \cite{Gubser:2000ec,Gubser:2000mm,Reall:2001ag}. For other cases, both $C_s^2$ and $C_v$ are singled value. The peak of $C_v$ becomes flattened with the increasing number of flavors.

\begin{figure}
    \centering
    \includegraphics[width=8cm]{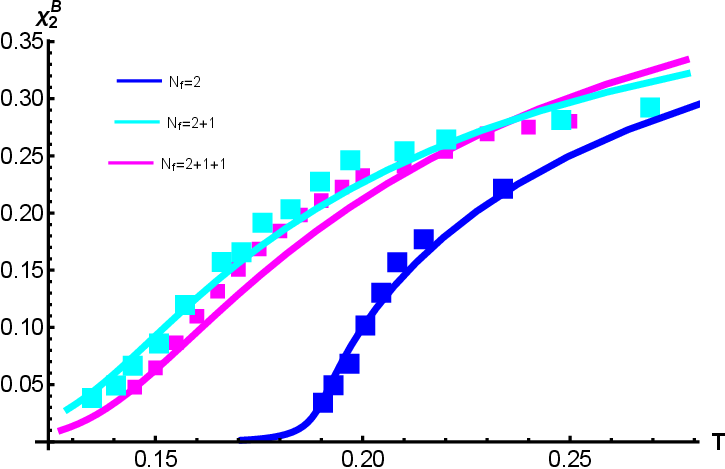}
    \caption{\label{mu0chi} The baryon number susceptibility as a function of temperature for different flavors. The squares represent lattice results from Ref. \cite{Datta:2016ukp}, Ref. \cite{Bazavov:2017dus}, and Ref. \cite{Bellwied:2015lba} for 2, 2+1, and 2+1+1 flavor, respectively. The unit of temperature is in GeV. }
\end{figure}

The comparison of our model with the lattice results for the baryon number susceptibility in different systems is shown in Fig.~\ref{mu0chi}. It is worth mentioning that the charm number susceptibility is very small and can be neglected around the phase transition temperature, as seen in Ref.~\cite{Bellwied:2015lba}. We only focus on the fitting range around the phase transition temperature since the QGP is strongly coupled in this case.

Unlike the potential reconstruction method described above, a numerical approach that constructs a family of five-dimensional black holes was investigated in the original papers \cite{Gubser:2008ny, DeWolfe:2010he}, and subsequently studied in Refs.~\cite{Critelli:2017oub, Grefa:2021qvt}. More recently, a similar model has been further developed in Refs.~\cite{He:2022amv, Zhao:2022uxc, Li:2023mpv, Zhao:2023gur}, incorporating lattice fitting. When comparing those models, the advantage of our model is that it has an analytical solution. The disadvantage is that we still make some assumptions about the form of $A(z)$ and $f(z)$. We plan to use machine learning to extract the numerical solutions of $A(z)$ and $f(z)$ without any assumptions in our future work.

\section{Thermodynamics for different flavors at finite chemical potential}\label{sec:05}
\subsection{Thermodynamics of 2-flavor system at finite chemical potential}
\begin{figure}
    \centering
    \includegraphics[width=16cm]{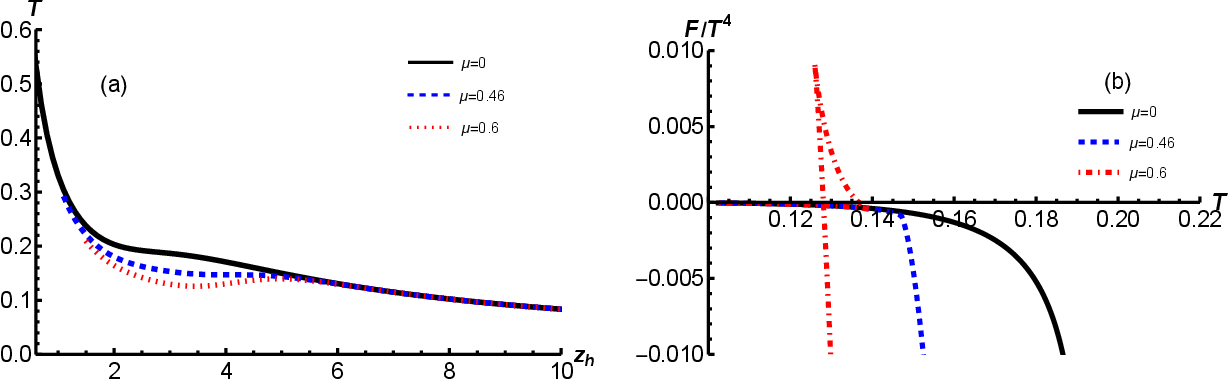}
    \caption{\label{2T}  (a) The temperature as a function of $z_h$ for 2 flavor at finite chemical potential. (b) The free energy as a function of $T$ for 2-flavor at chemical potential. The unit of the temperature and chemical potential are GeV. The unit of $z_h$ is in $\rm GeV^{-1}.$}
\end{figure}

In this section, we consider the 2-flavor system at finite chemical potential with fixed parameters determined by machine learning. At small chemical potential, the temperature is a single-valued function of $z_h$ which differs from the case of pure gluon as shown in Fig. \ref{2T} (a). As the chemical potential increases, the temperature becomes a multi-valued function of $z_h$, as depicted in Fig. \ref{2T} (b). The free energy is a single-valued function of the temperature at small chemical potential. However, at $\mu \geq 0.46 \, \rm GeV$, the free energy exhibits a swallowtail shape, indicating a first-order phase transition.

\begin{figure}
    \centering
    \includegraphics[width=16cm]{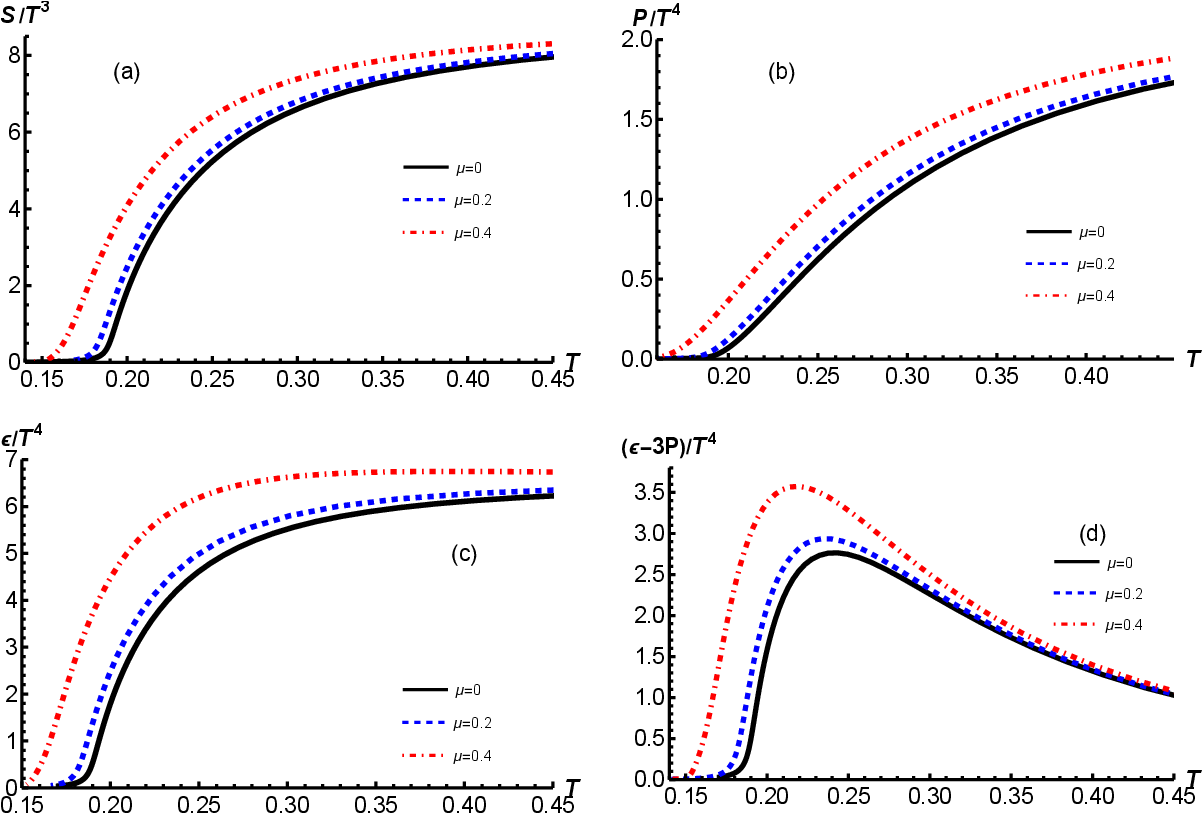}
    \caption{\label{2eos} (a) The entropy as a function of temperature for different chemical potentials in the 2-flavor system. (b) The pressure as a function of temperature for different chemical potentials in the 2-flavor system. (c) The energy as a function of temperature for different chemical potentials in the 2-flavor system. (d) The trace anomaly as a function of temperature for different chemical potentials in the 2-flavor system. The units of temperature and chemical potential are in GeV.}
\end{figure}

In Fig. \ref{2eos} (a), we present the entropy as a function of temperature at finite chemical potential. It is observed that the presence of a chemical potential increases the entropy value. However, at high temperatures, the results converge to a constant value. Similarly, the pressure, energy, and trace anomaly can be depicted as functions of temperature at various chemical potentials. The conclusions drawn for these quantities are similar to those for the entropy.

\begin{figure}
    \centering
    \includegraphics[width=16cm]{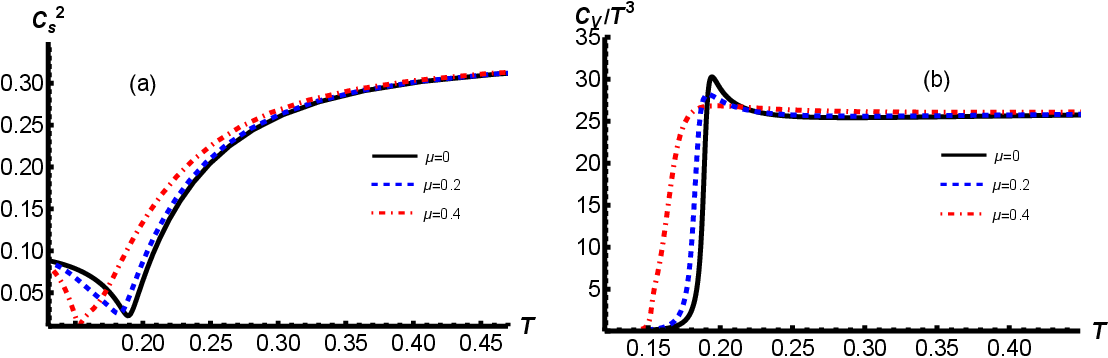}
    \caption{\label{2cscv} (a) The square of the speed of sound as a function of temperature for different chemical potentials in the 2-flavor system. (b) The specific heat as a function of temperature for different chemical potentials in the 2-flavor system. The units of temperature and chemical potential are in GeV.}
\end{figure}

Finally, the square of the speed of sound and the specific heat are depicted in Fig. \ref{2cscv}. The crossover does not constitute a real phase transition, which makes defining the transition temperature challenging. In this paper, we select the minimum of the speed of sound as the criterion for defining the transition temperature at vanishing chemical potential. The critical temperature for the 2-flavor system is determined to be $T_c = 0.189$ GeV. Unlike in the case of a pure gluon system, the specific heat here is always positive. The peak of the specific heat becomes flat with the increase of the chemical potential.

\subsection{Thermodynamics of 2+1-flavor system at finite chemical potential}
\begin{figure}
    \centering
    \includegraphics[width=16cm]{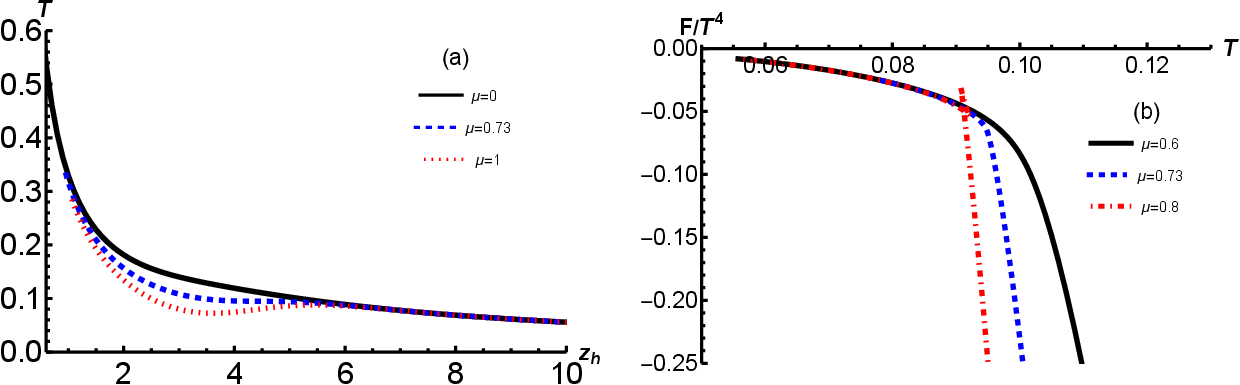}
    \caption{\label{3T}  (a) The temperature as a function of $z_h$ for different chemical potentials in the 2+1-flavor system. (b) The free energy as a function of $T$ for different chemical potentials in the 2+1-flavor system. The unit of the temperature and chemical potential is GeV. The unit of $z_h$ is in $\rm GeV^{-1}$. }
\end{figure}

We will turn to the 2+1 flavor in this section. The same procedure used to train the parameters and $k$ is turned on from this section. The temperature as a function of $z_h$ is shown in Fig. \ref{3T}. It also shows the temperature is a single-valued function of $z_h$ at small chemical potential and a multi-valued function of $z_h$ at large chemical potential. The free energy exhibits the same characteristics as in the case of the 2-flavor system.

\begin{figure}
    \centering
    \includegraphics[width=16cm]{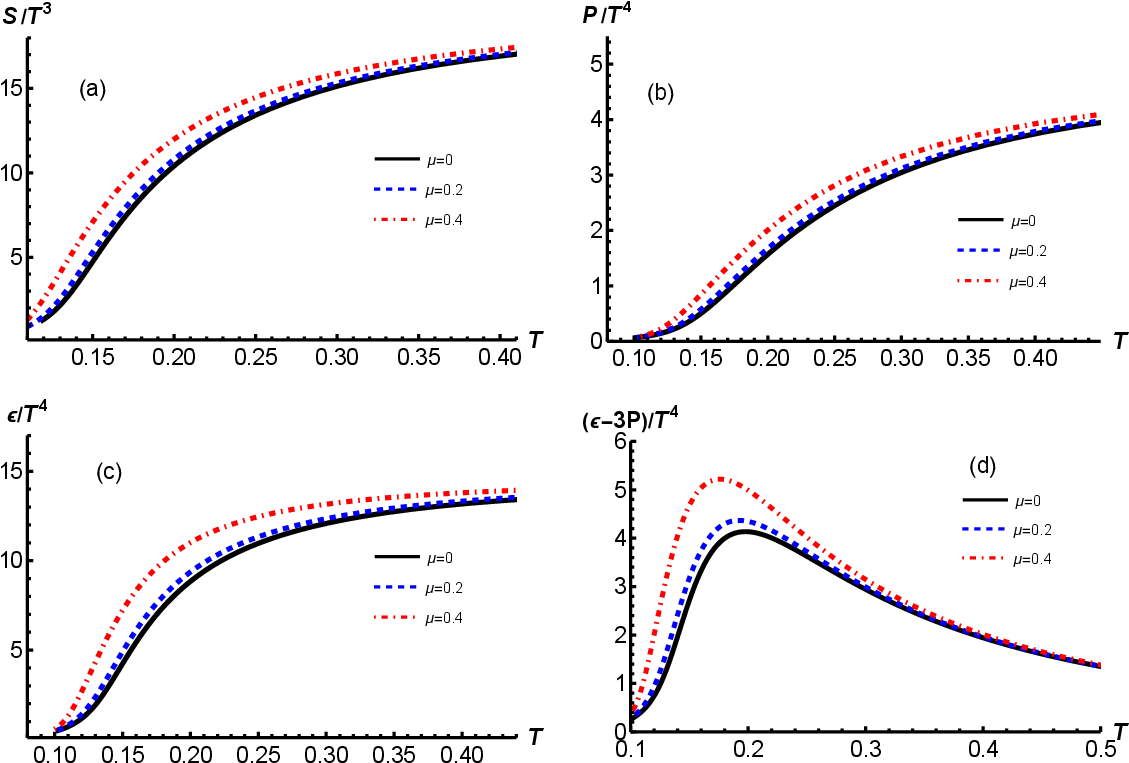}
    \caption{\label{3eos} (a) The entropy as a function of temperature for different chemical potentials in the 2+1-flavor system. (b) The pressure as a function of temperature for different chemical potentials in the 2+1-flavor system. (c) The energy as a function of temperature for different chemical potentials in the 2+1-flavor system. (d) The trace anomaly as a function of temperature for different chemical potentials in the 2+1-flavor system. The units of temperature and chemical potential are in GeV.}
\end{figure}

In addition, the EoS is also shown in Fig. \ref{3eos}. The conclusions drawn are analogous to the case of 2 flavor. At finite chemical potential, the entropy, pressure, energy, and trace anomaly are enhanced. At high temperatures, the differences in the EoS become small for different chemical potentials. However, compared with 2 flavor, we can see that the influence of chemical potential on the EoS is smaller.

\begin{figure}
    \centering
    \includegraphics[width=16cm]{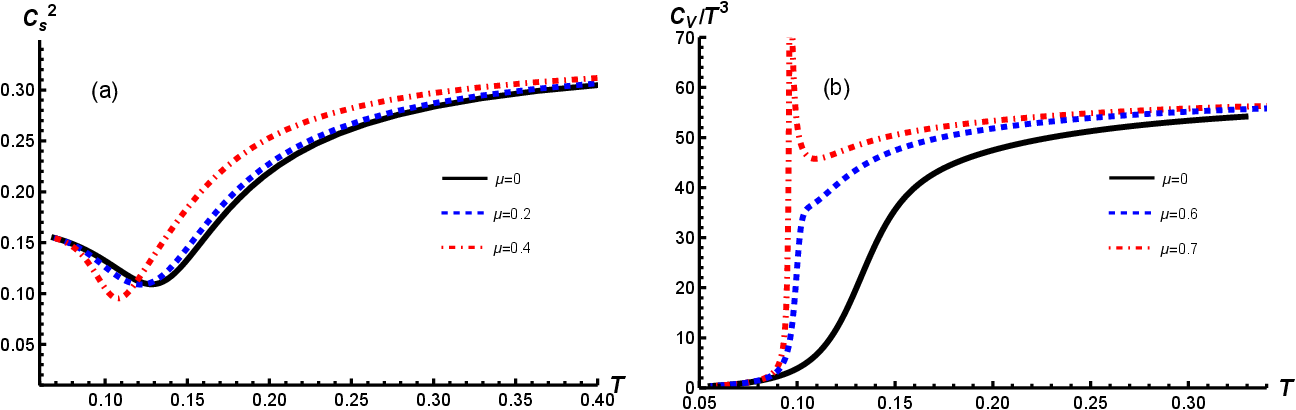}
    \caption{\label{3cscv} (a) The square of the speed of sound as a function for different chemical potentials in the 2+1-flavor system. (b) The specific heat as a function of temperature for different chemical potentials in the 2+1-flavor system. The units of temperature and chemical potential are in GeV.}
\end{figure}

The speed of sound and the specific heat are shown in Fig.~\ref{3cscv} for the 2+1 flavor. The crossover temperature is extracted for $T_c = 0.128~\rm GeV$ from the minimum of the speed of sound at vanishing chemical potential. The result is very close to the lattice $T_c = 0.132^{+3}_{-6}~\rm GeV$ of chiral phase transition \cite{HotQCD:2019xnw}. It is worth noting that the specific heat is different from the previous case. There is no peak at vanishing chemical potential. With the increase of chemical potential, the specific heat will show a peak as before. The peak becomes sharp with the increase of chemical potential.

\subsection{Thermodynamics of 2+1+1-flavor system at finite chemical potential}
\begin{figure}
    \centering
    \includegraphics[width=16cm]{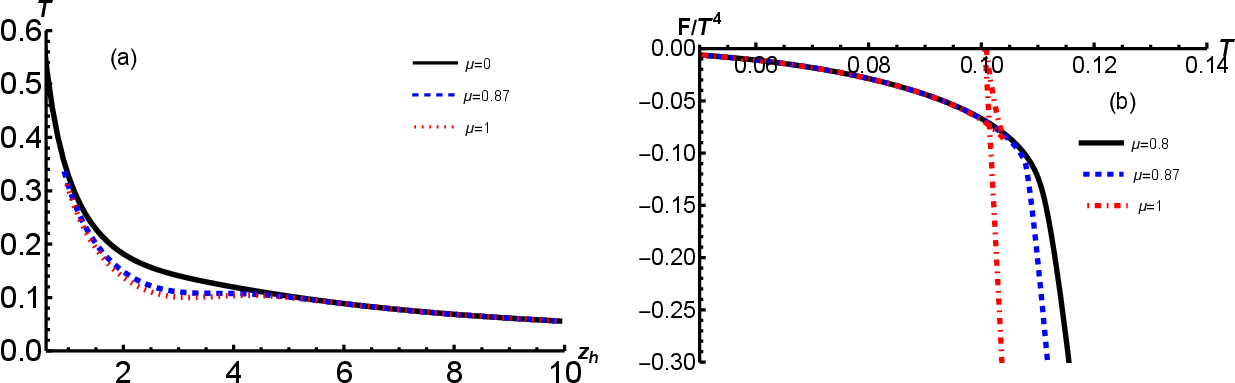}
    \caption{\label{4T} (a) The temperature as a function of the horizon for different chemical potentials in the 2+1+1-flavor system. (b) The free energy as a function of the temperature for different chemical potentials in the 2+1+1-flavor system. The units of the temperature and chemical potential are GeV. The unit of $z_h$ is in $\rm GeV^{-1}$. }
\end{figure}

In this section, thermodynamics of 2+1+1-system is discussed. The temperature as a function of $z_h$ is shown in Fig. \ref{4T} (a). It indicates that the temperature is a multi-valued function of $z_h$ at the chemical potential above 0.78 GeV. The free energy is also depicted in Fig. \ref{4T} (b), exhibiting a similar pattern to the previous case.

\begin{figure}
    \centering
    \includegraphics[width=16cm]{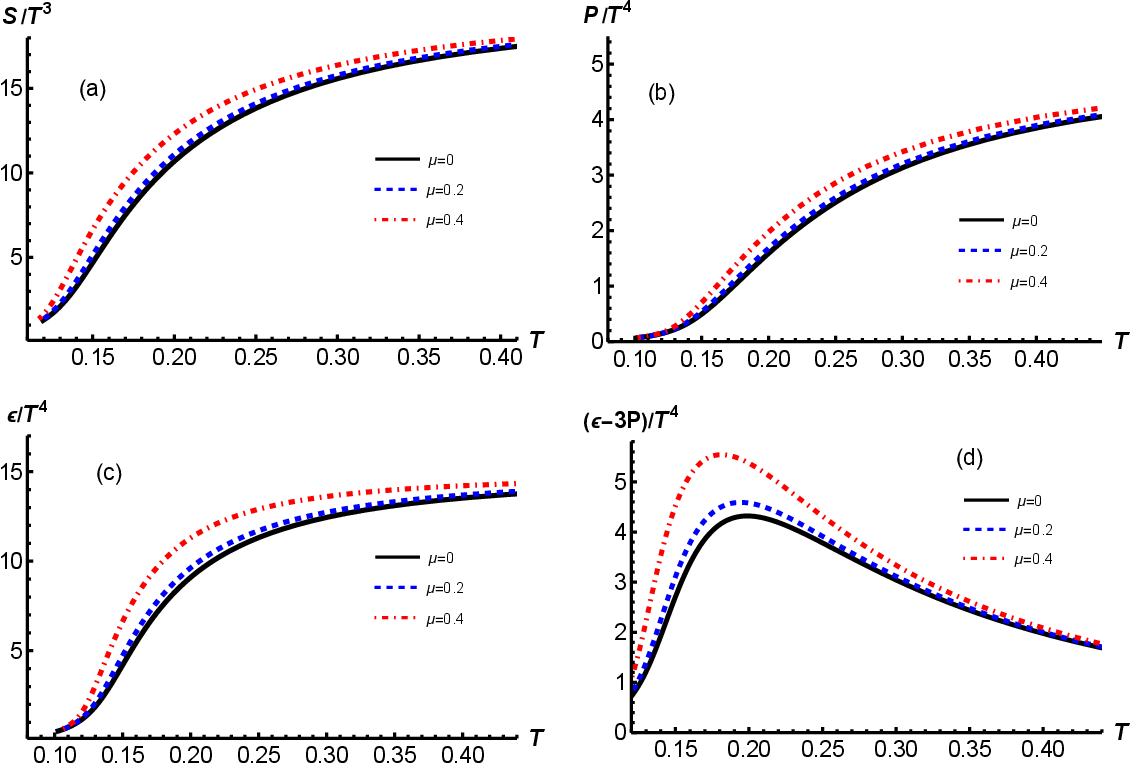}
    \caption{\label{4eos} (a) The entropy as a function of temperature for different chemical potentials in the 2+1+1-flavor system. (b) The pressure as a function of temperature for different chemical potentials in the 2+1+1-flavor system. (c) The energy as a function of temperature for different chemical potentials in the 2+1+1-flavor system. (d) The trace anomaly as a function of temperature for different chemical potentials in the 2+1+1-flavor system. The units of temperature and chemical potential are in GeV.}
\end{figure}

In addition, the EoS for 2+1+1-flavor at finite chemical potentials is also depicted in Fig. \ref{4eos}. The entropy, pressure, energy, and trace anomaly are enhanced at finite chemical potential, similar to the previous case. However, it has been found that systems with 2 flavors are more sensitive to changes in the chemical potential. The enhancement for systems with 2+1 and 2+1+1 flavors is lower compared to that of the 2-flavor system.

\begin{figure}
    \centering
    \includegraphics[width=16cm]{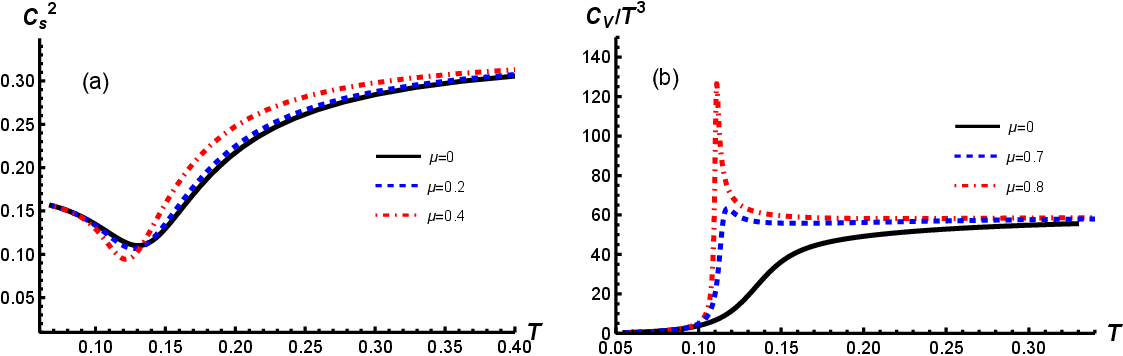}
    \caption{\label{4cscv} (a) The square of the speed of sound as a function of temperature for different chemical potentials in the 2+1+1-flavor system. (b) The specific heat as a function of temperature for different chemical potentials in the 2+1+1-flavor system. The units of temperature and chemical potential are in GeV.}
\end{figure}

The speed of sound and the specific heat are shown in Fig. \ref{4cscv} for the 2+1+1-flavor. With the increase in chemical potential, the minimum of $C_s^2$ will shift to the left. The sharp peak of $C_V$ occurs at a large chemical potential, similar to the 2+1-flavor case and distinct from the 2-flavor case.

\section{High-order baryon number susceptibilities}\label{sec:06}
\begin{figure}
    \centering
    \includegraphics[width=16cm]{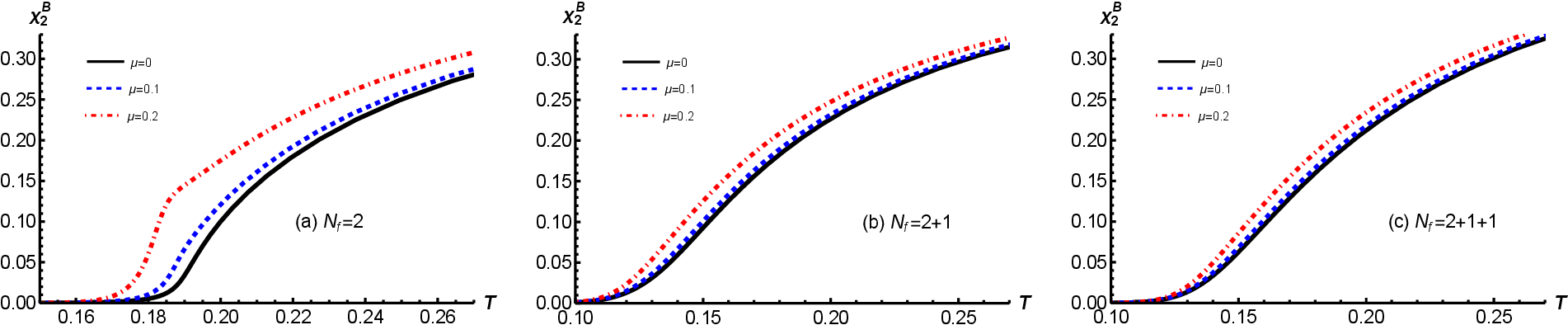}
    \caption{\label{chi2compare} (a) $\chi_2$ as a function of temperature at finite chemical potential for $N_f = 2$. (b) $\chi_2$ as a function of temperature at finite chemical potential for $N_f = 2 + 1$. (c) $\chi_2$ as a function of temperature at finite chemical potential for $N_f = 2 + 1 + 1$. The unit of $T$ is in $\rm GeV$.}
\end{figure}

We will discuss $\chi_2$ at finite temperature and chemical potential, as well as $\chi_4$ and $\chi_6$ at vanishing chemical potential, in this section. The $\chi_2$ is a different criterions for the crossover. The maximum slope of $\chi_2$ for the different chemical potential is often higher than $c_s^2$ \cite{Rougemont:2023gfz,Cai:2022omk}. Thus, the crossover temperature will be higher judging by $\chi_2$. We focus on the criterion of $c_s^2$ for the crossover in this paper. We draw the picture for $\chi_2$ as a function of $T$ for different values of $\mu$ in Fig. \ref{chi2compare}.

\begin{figure}
    \centering
    \includegraphics[width=16cm]{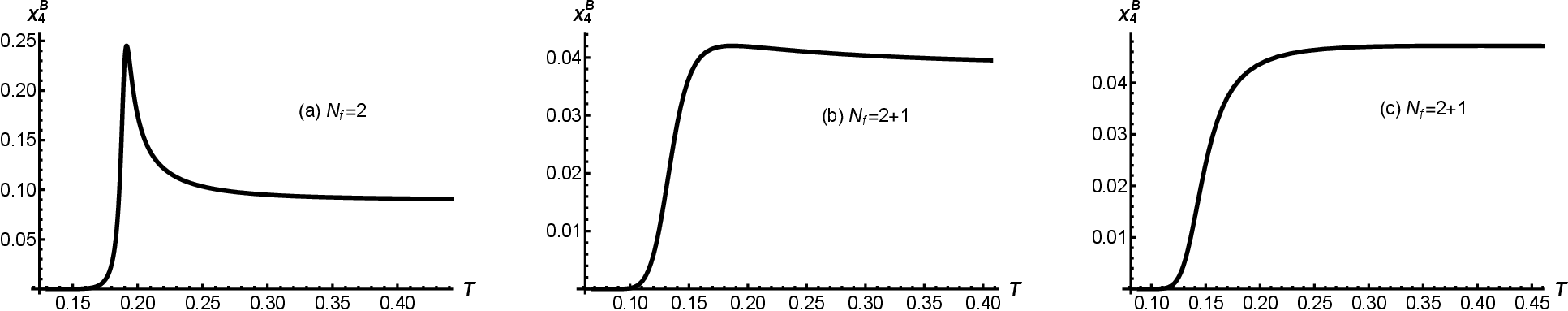}
    \caption{\label{chi4} (a) $\chi_4$ as a function of temperature for $N_f = 2$. (b) $\chi_4$ as a function of temperature for $N_f = 2 +1$. (c) $\chi_4$ as a function of temperature for $N_f = 2 +1+1$. The unit of $T$ is in $\rm GeV$.}
\end{figure}

\begin{figure}
    \centering
    \includegraphics[width=16cm]{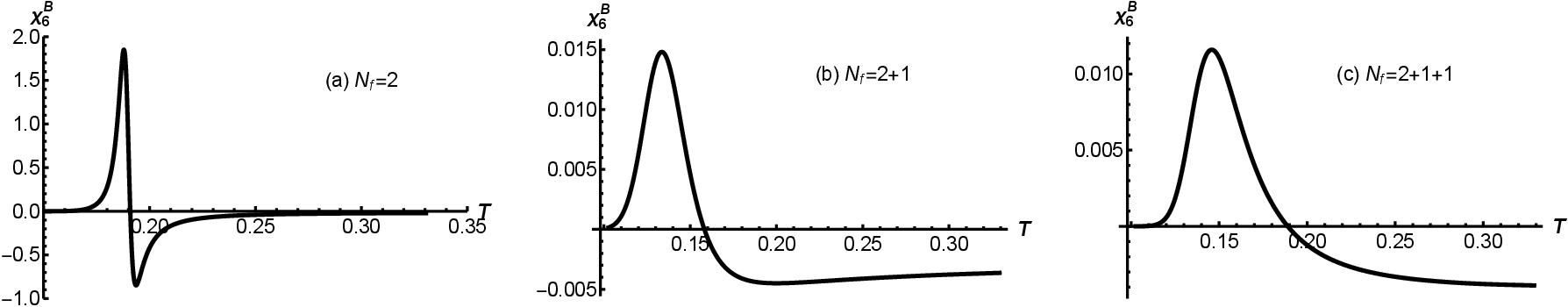}
    \caption{\label{chi6} (a) $\chi_6$ as a function of temperature for $N_f = 2$. (b) $\chi_6$ as a function of temperature for $N_f = 2 +1$. (c) $\chi_6$ as a function of temperature for $N_f = 2 +1+1$. The unit of $T$ is in $\rm GeV$.}
\end{figure}

Further, we have calculated the $\chi_4$ and $\chi_6$ in our model as shown in Fig.~\ref{chi4} and Fig.~\ref{chi6}. The qualitative behavior matches that of lattice QCD \cite{Datta:2016ukp,Bazavov:2020bjn,Borsanyi:2018grb}. We observe a trend: as the number of flavors increases, the peaks of both $\chi_4$ and $\chi_6$ tend to flatten. We aim to develop precise quantitative results in our future refined model. Higher-order baryon number susceptibilities, such as $\chi_8$, $\chi_{10}$, and $\chi_{12}$, will be investigated in future work.

\section{QCD Phase diagram for different flavors}\label{sec:07}
\begin{figure}
    \centering
    \includegraphics[width=16cm]{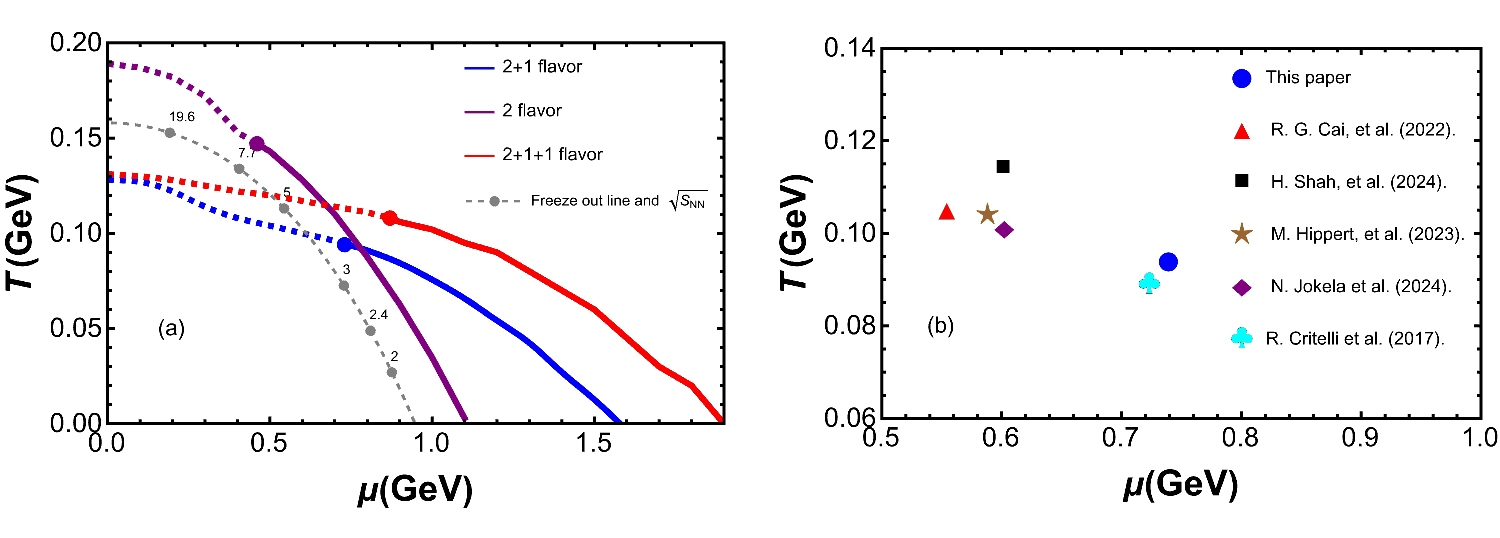}
    \caption{\label{cep} QCD phase diagram and the location of CEP predicted for different flavors in our model. The grey line represents the freeze out line \cite{Luo:2017faz}. The brown star is taken from Ref. \cite{Hippert:2023bel}, the clubs from Ref. \cite{Critelli:2017oub}, the black square from Ref. \cite{Shah:2024img}, the red triangle from Ref. \cite{Cai:2022omk}, and the purple diamond from Ref. \cite{Jokela:2024xgz}.}
\end{figure}

In this section, we summarize the locations of the CEP predicted by our model for all flavors, as shown in Fig. \ref{cep} (a). The crossover is determined from the minimum of the speed of sound at small chemical potentials. The free energy shows a multi-valued function of temperature at large chemical potential, which indicates the phase transition is first order. The critical temperature can be determined from the crossing point of the free energy curves for the first order. It is found that the CEP and the phase transition line of first order are above the freeze out line \cite{Luo:2017faz}. This model predicts the position of the CEP for the 2+1-flavor which is close to the recent results of other models \cite{Gao:2020qsj,Fu:2019hdw,Li:2018ygx}. The CEP in the $T - \mu$ plane locates at ($\mu_B^c$ = 0.46 GeV, $T^c$ = 0.147 GeV), ($\mu_B^c$ = 0.74 GeV, $T^c$ = 0.094 GeV), and ($\mu_B^c$ = 0.87 GeV, $T^c$ = 0.108 GeV) for the 2-flavor, 2+1-flavor systems, and 2+1+1-flavor systems, respectively. At zero temperature $T=0$, the critical chemical potential is found to be $\mu_B$ = 1.1 GeV, 1.6 GeV, 1.9 GeV for the 2-flavor, 2+1-flavor and 2+1+1-flavor systems, respectively, which implies that the effective Fermi momentum increases with increasing flavors. Therefore, this result confirms the qualitative correctness of the flavor-dependent model. In Fig. \ref{cep} (b), we compare our result with other holographic model results in Refs. \cite{Jokela:2024xgz,Critelli:2017oub,Shah:2024img,Hippert:2023bel,Cai:2022omk}, it is observed that recent holographic QCD models predict the location of CEP for 2+1 flavor is around $(T = 0.09\rm GeV \sim 0.114 \, \rm GeV, \mu = 0.555 \rm GeV \sim 0.74 \, \rm GeV$).

\section{Conclusion and outlook}\label{sec:08}
In this work, we input the information of EoS and baryon number susceptibility from lattice results to constrain the functions of $A(z)$ and $f(z)$ with the aid of machine learning in our model. Then, we provide predictions for the critical temperature of the deconfinement phase transition at vanishing chemical potential and the positions of the CEP at finite chemical potential for different systems.

It has been observed that: 1) At vanishing chemical potential $\mu=0$, the deconfinement phase transition is of first order for the pure gluon system with critical temperature $T_c = 0.265~\rm GeV$. The phase transition for the 2-flavor, 2+1-flavor systems, and 2+1+1-flavor system are crossover. The critical temperature decreases with increasing of flavors, and the difference between the 2+1-flavor and 2+1+1-flavor systems is invisible, which can be understood because the charm quark is heavy and its contribution to thermodynamic properties can be neglected.
2) At high chemical potential, the phase transition for the 2-flavor, 2+1-flavor systems, and 2+1+1-flavor systems are of first-order, and the critical endpoint (CEP) in the $T - \mu$ plane locates at ($\mu_B^c$=0.46 GeV, $T^c$=0.147 GeV), ($\mu_B^c$ = 0.74 GeV, $T^c$ = 0.094 GeV), and ($\mu_B^c$= 0.87 GeV, $T^c$ = 0.108 GeV), respectively. The difference between the 2+1-flavor and 2+1+1-flavor systems increases with the chemical potential, and the location of CEP for 2+1+1-flavor system deviates explicitly from that of the 2+1-flavor system. Both 2+1-flavor and 2+1+1-flavor systems differ significantly from the 2-flavor system. Besides, at zero temperature $T=0$, the critical chemical potential is found to be $\mu_B$ = 1.1 GeV, 1.6 GeV, 1.9 GeV for the 2-flavor, 2+1-flavor and 2+1+1-flavor systems, respectively, which implies that the effective Fermi momentum increases with increasing flavors. Therefore, this result confirms the qualitative correctness of the flavor-dependent model.

This work represents an attempt to construct an analytical holographic model using machine learning. This analytical model can give different phase structures for different flavors. We hope that this model will be benefit for the searching of CEP in the QCD phase diagram and enhance our understanding of phase transition characteristics within the domain of strong interactions. Moreover, many other physical quantities, in addition to the drag force and the jet quenching parameter \cite{Chen:2024epd}, can be calculated in future work based on our model. For instance, we can use this model to calculate the internal structure of neutron stars. Further, we aim to incorporate more information into the holographic QCD and construct an even more realistic holographic model through machine learning.

\section*{Acknowledgments}
This work is supported in part by the National Natural Science Foundation of China (NSFC) Grant Nos: 12405154, 12235016, 12221005, and the Strategic Priority Research Program of Chinese Academy of Sciences under Grant No XDB34030000, and the Fundamental Research Funds for the Central Universities, Open Fund for Key Laboratories of the Ministry of Education under Grants No. QLPL2024P01.

\bibliographystyle{unsrt}
\bibliography{ref}

\begin{thebibliography}{100}

\bibitem{ALICE:2010suc}
K~Aamodt et~al.
\newblock {Elliptic flow of charged particles in Pb-Pb collisions at 2.76 TeV}.
\newblock {\em Phys. Rev. Lett.}, 105:252302, 2010.

\bibitem{ATLAS:2011ah}
Georges Aad et~al.
\newblock {Measurement of the pseudorapidity and transverse momentum dependence
  of the elliptic flow of charged particles in lead-lead collisions at
  $\sqrt{s_{NN}}=2.76$ TeV with the ATLAS detector}.
\newblock {\em Phys. Lett. B}, 707:330--348, 2012.

\bibitem{CMS:2012xss}
Serguei Chatrchyan et~al.
\newblock {Centrality Dependence of Dihadron Correlations and Azimuthal
  anisotropy Harmonics in PbPb Collisions at $\sqrt{s_{NN}}=2.76$ TeV}.
\newblock {\em Eur. Phys. J. C}, 72:2012, 2012.

\bibitem{Fukushima:2010bq}
Kenji Fukushima and Tetsuo Hatsuda.
\newblock {The phase diagram of dense QCD}.
\newblock {\em Rept. Prog. Phys.}, 74:014001, 2011.

\bibitem{deForcrand:2009zkb}
Philippe de~Forcrand.
\newblock {Simulating QCD at finite density}.
\newblock {\em PoS}, LAT2009:010, 2009.

\bibitem{Pisarski:1983ms}
Robert~D. Pisarski and Frank Wilczek.
\newblock {Remarks on the Chiral Phase Transition in Chromodynamics}.
\newblock {\em Phys. Rev. D}, 29:338--341, 1984.

\bibitem{Asakawa:1989bq}
M.~Asakawa and K.~Yazaki.
\newblock {Chiral Restoration at Finite Density and Temperature}.
\newblock {\em Nucl. Phys. A}, 504:668--684, 1989.

\bibitem{Stephanov:1998dy}
Misha~A. Stephanov, K.~Rajagopal, and Edward~V. Shuryak.
\newblock {Signatures of the tricritical point in QCD}.
\newblock {\em Phys. Rev. Lett.}, 81:4816--4819, 1998.

\bibitem{Hatta:2002sj}
Yoshitaka Hatta and Takashi Ikeda.
\newblock {Universality, the QCD critical / tricritical point and the quark
  number susceptibility}.
\newblock {\em Phys. Rev. D}, 67:014028, 2003.

\bibitem{Stephanov:1999zu}
Misha~A. Stephanov, K.~Rajagopal, and Edward~V. Shuryak.
\newblock {Event-by-event fluctuations in heavy ion collisions and the QCD
  critical point}.
\newblock {\em Phys. Rev. D}, 60:114028, 1999.

\bibitem{Hatta:2003wn}
Y.~Hatta and M.~A. Stephanov.
\newblock {Proton number fluctuation as a signal of the QCD critical endpoint}.
\newblock {\em Phys. Rev. Lett.}, 91:102003, 2003.
\newblock [Erratum: Phys.Rev.Lett. 91, 129901 (2003)].

\bibitem{Schwarz:1999dj}
T.~M. Schwarz, S.~P. Klevansky, and G.~Papp.
\newblock {The Phase diagram and bulk thermodynamical quantities in the NJL
  model at finite temperature and density}.
\newblock {\em Phys. Rev. C}, 60:055205, 1999.

\bibitem{Zhuang:2000ub}
P.~Zhuang, M.~Huang, and Z.~Yang.
\newblock {Density effect on hadronization of a quark plasma}.
\newblock {\em Phys. Rev. C}, 62:054901, 2000.

\bibitem{Chodos:1974je}
A.~Chodos, R.~L. Jaffe, K.~Johnson, Charles~B. Thorn, and V.~F. Weisskopf.
\newblock {A New Extended Model of Hadrons}.
\newblock {\em Phys. Rev. D}, 9:3471--3495, 1974.

\bibitem{DeRujula:1975qlm}
A.~De~Rujula, Howard Georgi, and S.~L. Glashow.
\newblock {Hadron Masses in a Gauge Theory}.
\newblock {\em Phys. Rev. D}, 12:147--162, 1975.

\bibitem{Nambu:1961tp}
Yoichiro Nambu and G.~Jona-Lasinio.
\newblock {Dynamical Model of Elementary Particles Based on an Analogy with
  Superconductivity. 1.}
\newblock {\em Phys. Rev.}, 122:345--358, 1961.

\bibitem{Nambu:1961fr}
Yoichiro Nambu and G.~Jona-Lasinio.
\newblock {Dynamical model of elementary particles based on an analogy with
  superconductivity. II.}
\newblock {\em Phys. Rev.}, 124:246--254, 1961.

\bibitem{Gao:2016qkh}
Fei Gao and Yu-xin Liu.
\newblock {QCD phase transitions via a refined truncation of Dyson-Schwinger
  equations}.
\newblock {\em Phys. Rev. D}, 94(7):076009, 2016.

\bibitem{Qin:2010nq}
Si-xue Qin, Lei Chang, Huan Chen, Yu-xin Liu, and Craig~D. Roberts.
\newblock {Phase diagram and critical endpoint for strongly-interacting
  quarks}.
\newblock {\em Phys. Rev. Lett.}, 106:172301, 2011.

\bibitem{Shi:2014zpa}
Chao Shi, Yong-Long Wang, Yu~Jiang, Zhu-Fang Cui, and Hong-Shi Zong.
\newblock {Locate QCD Critical End Point in a Continuum Model Study}.
\newblock {\em JHEP}, 07:014, 2014.

\bibitem{Fischer:2014ata}
Christian~S. Fischer, Jan Luecker, and Christian~A. Welzbacher.
\newblock {Phase structure of three and four flavor QCD}.
\newblock {\em Phys. Rev. D}, 90(3):034022, 2014.

\bibitem{McLerran:2008ua}
Larry McLerran, Krzystof Redlich, and Chihiro Sasaki.
\newblock {Quarkyonic Matter and Chiral Symmetry Breaking}.
\newblock {\em Nucl. Phys. A}, 824:86--100, 2009.

\bibitem{Sasaki:2010jz}
Takahiro Sasaki, Yuji Sakai, Hiroaki Kouno, and Masanobu Yahiro.
\newblock {QCD phase diagram at finite baryon and isospin chemical potentials}.
\newblock {\em Phys. Rev. D}, 82:116004, 2010.

\bibitem{Li:2018ygx}
Zhibin Li, Kun Xu, Xinyang Wang, and Mei Huang.
\newblock {The kurtosis of net baryon number fluctuations from a realistic
  Polyakov\textendash{}Nambu\textendash{}Jona-Lasinio model along the
  experimental freeze-out line}.
\newblock {\em Eur. Phys. J. C}, 79(3):245, 2019.

\bibitem{Sun:2023kuu}
Fei Sun, Kun Xu, and Mei Huang.
\newblock {Splitting of chiral and deconfinement phase transitions induced by
  rotation}.
\newblock {\em Phys. Rev. D}, 108(9):096007, 2023.

\bibitem{Bao:2024glw}
Yan-Ru Bao and Sheng-Qin Feng.
\newblock {Effects of tensor spin polarization on the chiral restoration and
  deconfinement phase transitions}.
\newblock 3 2024.

\bibitem{Fu:2019hdw}
Wei-jie Fu, Jan~M. Pawlowski, and Fabian Rennecke.
\newblock {QCD phase structure at finite temperature and density}.
\newblock {\em Phys. Rev. D}, 101(5):054032, 2020.

\bibitem{Zhang:2017icm}
Hui Zhang, Defu Hou, Toru Kojo, and Bin Qin.
\newblock {Functional renormalization group study of the quark-meson model with
  $\omega$ meson}.
\newblock {\em Phys. Rev. D}, 96(11):114029, 2017.

\bibitem{Becattini:2016xct}
F.~Becattini, J.~Steinheimer, R.~Stock, and M.~Bleicher.
\newblock {Hadronization conditions in relativistic nuclear collisions and the
  QCD pseudo-critical line}.
\newblock {\em Phys. Lett. B}, 764:241--246, 2017.

\bibitem{Fujimoto:2021xix}
Yuki Fujimoto, Kenji Fukushima, and Yoshimasa Hidaka.
\newblock {Deconfining Phase Boundary of Rapidly Rotating Hot and Dense Matter
  and Analysis of Moment of Inertia}.
\newblock {\em Phys. Lett. B}, 816:136184, 2021.

\bibitem{Maldacena:1997re}
Juan~Martin Maldacena.
\newblock {The Large N limit of superconformal field theories and
  supergravity}.
\newblock {\em Adv. Theor. Math. Phys.}, 2:231--252, 1998.

\bibitem{Erdmenger:2007cm}
Johanna Erdmenger, Nick Evans, Ingo Kirsch, and Ed~Threlfall.
\newblock {Mesons in Gauge/Gravity Duals - A Review}.
\newblock {\em Eur. Phys. J. A}, 35:81--133, 2008.

\bibitem{Brodsky:2014yha}
Stanley~J. Brodsky, Guy~F. de~Teramond, Hans~Gunter Dosch, and Joshua Erlich.
\newblock {Light-Front Holographic QCD and Emerging Confinement}.
\newblock {\em Phys. Rept.}, 584:1--105, 2015.

\bibitem{Casalderrey-Solana:2011dxg}
Jorge Casalderrey-Solana, Hong Liu, David Mateos, Krishna Rajagopal, and
  Urs~Achim Wiedemann.
\newblock {\em {Gauge/String Duality, Hot QCD and Heavy Ion Collisions}}.
\newblock Cambridge University Press, 2014.

\bibitem{Adams:2012th}
Allan Adams, Lincoln~D. Carr, Thomas Sch\"afer, Peter Steinberg, and John~E.
  Thomas.
\newblock {Strongly Correlated Quantum Fluids: Ultracold Quantum Gases, Quantum
  Chromodynamic Plasmas, and Holographic Duality}.
\newblock {\em New J. Phys.}, 14:115009, 2012.

\bibitem{Rougemont:2023gfz}
Romulo Rougemont, Joaquin Grefa, Mauricio Hippert, Jorge Noronha, Jacquelyn
  Noronha-Hostler, Israel Portillo, and Claudia Ratti.
\newblock {Hot QCD Phase Diagram From Holographic Einstein-Maxwell-Dilaton
  Models}.
\newblock 7 2023.

\bibitem{Jarvinen:2022doa}
Matti Jarvinen.
\newblock {Holographic dense QCD in the Veneziano limit}.
\newblock {\em EPJ Web Conf.}, 274:08006, 2022.

\bibitem{Gubser:2008ny}
Steven~S. Gubser and Abhinav Nellore.
\newblock {Mimicking the QCD equation of state with a dual black hole}.
\newblock {\em Phys. Rev. D}, 78:086007, 2008.

\bibitem{DeWolfe:2010he}
Oliver DeWolfe, Steven~S. Gubser, and Christopher Rosen.
\newblock {A holographic critical point}.
\newblock {\em Phys. Rev. D}, 83:086005, 2011.

\bibitem{He:2013qq}
Song He, Shang-Yu Wu, Yi~Yang, and Pei-Hung Yuan.
\newblock {Phase Structure in a Dynamical Soft-Wall Holographic QCD Model}.
\newblock {\em JHEP}, 04:093, 2013.

\bibitem{Yang:2014bqa}
Yi~Yang and Pei-Hung Yuan.
\newblock {A Refined Holographic QCD Model and QCD Phase Structure}.
\newblock {\em JHEP}, 11:149, 2014.

\bibitem{Yang:2015aia}
Yi~Yang and Pei-Hung Yuan.
\newblock {Confinement-deconfinement phase transition for heavy quarks in a
  soft wall holographic QCD model}.
\newblock {\em JHEP}, 12:161, 2015.

\bibitem{Dudal:2017max}
David Dudal and Subhash Mahapatra.
\newblock {Thermal entropy of a quark-antiquark pair above and below
  deconfinement from a dynamical holographic QCD model}.
\newblock {\em Phys. Rev. D}, 96(12):126010, 2017.

\bibitem{Dudal:2018ztm}
David Dudal and Subhash Mahapatra.
\newblock {Interplay between the holographic QCD phase diagram and entanglement
  entropy}.
\newblock {\em JHEP}, 07:120, 2018.

\bibitem{Fang:2015ytf}
Zhen Fang, Song He, and Danning Li.
\newblock {Chiral and Deconfining Phase Transitions from Holographic QCD
  Study}.
\newblock {\em Nucl. Phys. B}, 907:187--207, 2016.

\bibitem{Li:2022erd}
Ying-Ying Li, Xing-Lin Liu, Xin-Yi Liu, and Zhen Fang.
\newblock {Correlations between the deconfining and chiral transitions in
  holographic QCD}.
\newblock {\em Phys. Rev. D}, 105(3):034019, 2022.

\bibitem{Critelli:2017oub}
Renato Critelli, Jorge Noronha, Jacquelyn Noronha-Hostler, Israel Portillo,
  Claudia Ratti, and Romulo Rougemont.
\newblock {Critical point in the phase diagram of primordial quark-gluon matter
  from black hole physics}.
\newblock {\em Phys. Rev. D}, 96(9):096026, 2017.

\bibitem{Grefa:2021qvt}
Joaquin Grefa, Jorge Noronha, Jacquelyn Noronha-Hostler, Israel Portillo,
  Claudia Ratti, and Romulo Rougemont.
\newblock {Hot and dense quark-gluon plasma thermodynamics from holographic
  black holes}.
\newblock {\em Phys. Rev. D}, 104(3):034002, 2021.

\bibitem{Arefeva:2020vae}
Irina~Ya. Aref'eva, Kristina Rannu, and Pavel Slepov.
\newblock {Holographic model for heavy quarks in anisotropic hot dense QGP with
  external magnetic field}.
\newblock {\em JHEP}, 07:161, 2021.

\bibitem{Chen:2018vty}
Xun Chen, Danning Li, and Mei Huang.
\newblock {Criticality of QCD in a holographic QCD model with critical end
  point}.
\newblock {\em Chin. Phys. C}, 43(2):023105, 2019.

\bibitem{Chen:2020ath}
Xun Chen, Lin Zhang, Danning Li, Defu Hou, and Mei Huang.
\newblock {Gluodynamics and deconfinement phase transition under rotation from
  holography}.
\newblock {\em JHEP}, 07:132, 2021.

\bibitem{Zhou:2020ssi}
Jing Zhou, Xun Chen, Yan-Qing Zhao, and Jialun Ping.
\newblock {Thermodynamics of heavy quarkonium in a magnetic field background}.
\newblock {\em Phys. Rev. D}, 102(8):086020, 2020.

\bibitem{Chen:2019rez}
Xun Chen, Danning Li, Defu Hou, and Mei Huang.
\newblock {Quarkyonic phase from quenched dynamical holographic QCD model}.
\newblock {\em JHEP}, 03:073, 2020.

\bibitem{Wang:2024szr}
Jia-Hao Wang and Sheng-Qin Feng.
\newblock {Rotation effect on the deconfinement phase transition in holographic
  QCD}.
\newblock {\em Phys. Rev. D}, 109(6):066019, 2024.

\bibitem{He:2022amv}
Song He, Li~Li, Zhibin Li, and Shao-Jiang Wang.
\newblock {Gravitational waves and primordial black hole productions from
  gluodynamics by holography}.
\newblock {\em Sci. China Phys. Mech. Astron.}, 67(4):240411, 2024.

\bibitem{Fu:2024wkn}
Qingxuan Fu, Song He, Li~Li, and Zhibin Li.
\newblock {Revisiting holographic model for thermal and dense QCD with a
  critical point}.
\newblock 4 2024.

\bibitem{Liu:2023pbt}
Xin-Yi Liu, Xiao-Chang Peng, Yue-Liang Wu, and Zhen Fang.
\newblock {A holographic study on QCD phase transition and phase diagram with
  two flavors}.
\newblock 12 2023.

\bibitem{Liu:2024efy}
Xin-Yi Liu, Yue-Liang Wu, and Zhen Fang.
\newblock {A holographic study on QCD phase transition and neutron star
  properties}.
\newblock 12 2024.

\bibitem{Cao:2024jgt}
Xuanmin Cao and Hui Liu.
\newblock {The impact of the phase transition on Quark-Gluon Plasma with an
  extremely strong magnetic field in holographic QCD}.
\newblock 8 2024.

\bibitem{Jokela:2024xgz}
Niko Jokela, Matti J\"arvinen, and Aleksi Piispa.
\newblock {Refining holographic models of the quark-gluon plasma}.
\newblock 5 2024.

\bibitem{Zhou:2023pti}
Kai Zhou, Lingxiao Wang, Long-Gang Pang, and Shuzhe Shi.
\newblock {Exploring QCD matter in extreme conditions with Machine Learning}.
\newblock {\em Prog. Part. Nucl. Phys.}, 104084:2023, 3 2023.

\bibitem{Hashimoto:2018ftp}
Koji Hashimoto, Sotaro Sugishita, Akinori Tanaka, and Akio Tomiya.
\newblock {Deep learning and the AdS/CFT correspondence}.
\newblock {\em Phys. Rev. D}, 98(4):046019, 2018.

\bibitem{Akutagawa:2020yeo}
Tetsuya Akutagawa, Koji Hashimoto, and Takayuki Sumimoto.
\newblock {Deep Learning and AdS/QCD}.
\newblock {\em Phys. Rev. D}, 102(2):026020, 2020.

\bibitem{Hashimoto:2018bnb}
Koji Hashimoto, Sotaro Sugishita, Akinori Tanaka, and Akio Tomiya.
\newblock {Deep Learning and Holographic QCD}.
\newblock {\em Phys. Rev. D}, 98(10):106014, 2018.

\bibitem{Yan:2020wcd}
Yu-Kun Yan, Shao-Feng Wu, Xian-Hui Ge, and Yu~Tian.
\newblock {Deep learning black hole metrics from shear viscosity}.
\newblock {\em Phys. Rev. D}, 102(10):101902, 4 2020.

\bibitem{Hashimoto:2021ihd}
Koji Hashimoto, Keisuke Ohashi, and Takayuki Sumimoto.
\newblock {Deriving the dilaton potential in improved holographic QCD from the
  meson spectrum}.
\newblock {\em Phys. Rev. D}, 105(10):106008, 2022.

\bibitem{Song:2020agw}
Mugeon Song, Maverick S.~H. Oh, Yongjun Ahn, and Keun-Young Kima.
\newblock {AdS/Deep-Learning made easy: simple examples}.
\newblock {\em Chin. Phys. C}, 45(7):073111, 2021.

\bibitem{Chang:2024ksq}
Wen-Bin Chang and De-fu Hou.
\newblock {Heavy quarkonium spectral function in an anisotropic background}.
\newblock {\em Phys. Rev. D}, 109(8):086010, 2024.

\bibitem{Ahn:2024gjf}
Byoungjoon Ahn, Hyun-Sik Jeong, Keun-Young Kim, and Kwan Yun.
\newblock {Deep learning bulk spacetime from boundary optical conductivity}.
\newblock {\em JHEP}, 03:141, 2024.

\bibitem{Gu:2024lrz}
Zhuo-Fan Gu, Yu-Kun Yan, and Shao-Feng Wu.
\newblock {Neural ODEs for holographic transport models without translation
  symmetry}.
\newblock 1 2024.

\bibitem{Li:2022zjc}
Kai Li, Yi~Ling, Peng Liu, and Meng-He Wu.
\newblock {Learning the black hole metric from holographic conductivity}.
\newblock {\em Phys. Rev. D}, 107(6):066021, 2023.

\bibitem{Chen:2024ckb}
Xun Chen and Mei Huang.
\newblock {Machine learning holographic black hole from lattice QCD equation of
  state}.
\newblock {\em Phys. Rev. D}, 109(5):L051902, 2024.

\bibitem{Zhang:2022uin}
Lin Zhang and Mei Huang.
\newblock {Holographic cold dense matter constrained by neutron stars}.
\newblock {\em Phys. Rev. D}, 106(9):096028, 2022.

\bibitem{Li:2020hau}
Meng-Wei Li, Yi~Yang, and Pei-Hung Yuan.
\newblock {Analytic Study on Chiral Phase Transition in Holographic QCD}.
\newblock {\em JHEP}, 02:055, 2021.

\bibitem{Yang:2017oer}
Yi~Yang and Pei-Hung Yuan.
\newblock {Universal Behaviors of Speed of Sound from Holography}.
\newblock {\em Phys. Rev. D}, 97(12):126009, 2018.

\bibitem{Gursoy:2017wzz}
Umut Gursoy, Matti Jarvinen, and Govert Nijs.
\newblock {Holographic QCD in the Veneziano Limit at a Finite Magnetic Field
  and Chemical Potential}.
\newblock {\em Phys. Rev. Lett.}, 120(24):242002, 2018.

\bibitem{Floerchinger:2015efa}
Stefan Floerchinger and Mauricio Martinez.
\newblock {Fluid dynamic propagation of initial baryon number perturbations on
  a Bjorken flow background}.
\newblock {\em Phys. Rev. C}, 92(6):064906, 2015.

\bibitem{Li:2023mpv}
Zhibin Li, Jingmin Liang, Song He, and Li~Li.
\newblock {Holographic study of higher-order baryon number susceptibilities at
  finite temperature and density}.
\newblock {\em Phys. Rev. D}, 108(4):046008, 2023.

\bibitem{HotQCD:2014kol}
A.~Bazavov et~al.
\newblock {Equation of state in ( 2+1 )-flavor QCD}.
\newblock {\em Phys. Rev. D}, 90:094503, 2014.

\bibitem{Bazavov:2017dus}
A.~Bazavov et~al.
\newblock {The QCD Equation of State to $\mathcal{O}(\mu_B^6)$ from Lattice
  QCD}.
\newblock {\em Phys. Rev. D}, 95(5):054504, 2017.

\bibitem{Borsanyi:2012ve}
Sz. Borsanyi, G.~Endrodi, Z.~Fodor, S.~D. Katz, and K.~K. Szabo.
\newblock {Precision SU(3) lattice thermodynamics for a large temperature
  range}.
\newblock {\em JHEP}, 07:056, 2012.

\bibitem{Burger:2014xga}
Florian Burger, Ernst-Michael Ilgenfritz, Maria~Paola Lombardo, and Michael
  M\"uller-Preussker.
\newblock {Equation of state of quark-gluon matter from lattice QCD with two
  flavors of twisted mass Wilson fermions}.
\newblock {\em Phys. Rev. D}, 91(7):074504, 2015.

\bibitem{Ratti:2013uta}
Claudia Ratti, Szabolcs Borsanyi, Gergely Endrodi, Zoltan Fodor, Sandor~D.
  Katz, Stefan Krieg, Chris Schroeder, and Kalman~K. Szabo.
\newblock {Lattice QCD thermodynamics in the presence of the charm quark}.
\newblock {\em Nucl. Phys. A}, 904-905:869c--872c, 2013.

\bibitem{He:2020fdi}
Song He, Yi~Yang, and Pei-Hung Yuan.
\newblock {Analytic Study of Magnetic Catalysis in Holographic QCD}.
\newblock 4 2020.

\bibitem{Gregory:1993vy}
R.~Gregory and R.~Laflamme.
\newblock {Black strings and p-branes are unstable}.
\newblock {\em Phys. Rev. Lett.}, 70:2837--2840, 1993.

\bibitem{Gregory:1994bj}
Ruth Gregory and Raymond Laflamme.
\newblock {The Instability of charged black strings and p-branes}.
\newblock {\em Nucl. Phys. B}, 428:399--434, 1994.

\bibitem{Gubser:2000ec}
Steven~S. Gubser and Indrajit Mitra.
\newblock {Instability of charged black holes in Anti-de Sitter space}.
\newblock {\em Clay Math. Proc.}, 1:221, 2002.

\bibitem{Gubser:2000mm}
Steven~S. Gubser and Indrajit Mitra.
\newblock {The Evolution of unstable black holes in anti-de Sitter space}.
\newblock {\em JHEP}, 08:018, 2001.

\bibitem{Reall:2001ag}
Harvey~S. Reall.
\newblock {Classical and thermodynamic stability of black branes}.
\newblock {\em Phys. Rev. D}, 64:044005, 2001.

\bibitem{Bellwied:2015lba}
R.~Bellwied, S.~Borsanyi, Z.~Fodor, S.~D. Katz, A.~Pasztor, C.~Ratti, and K.~K.
  Szabo.
\newblock {Fluctuations and correlations in high temperature QCD}.
\newblock {\em Phys. Rev. D}, 92(11):114505, 2015.

\bibitem{Datta:2016ukp}
Saumen Datta, Rajiv~V. Gavai, and Sourendu Gupta.
\newblock {Quark number susceptibilities and equation of state at finite
  chemical potential in staggered QCD with Nt=8}.
\newblock {\em Phys. Rev. D}, 95(5):054512, 2017.

\bibitem{Zhao:2022uxc}
Yan-Qing Zhao, Song He, Defu Hou, Li~Li, and Zhibin Li.
\newblock {Phase diagram of holographic thermal dense QCD matter with
  rotation}.
\newblock {\em JHEP}, 04:115, 2023.

\bibitem{Zhao:2023gur}
Yan-Qing Zhao, Song He, Defu Hou, Li~Li, and Zhibin Li.
\newblock {Phase structure and critical phenomena in two-flavor QCD by
  holography}.
\newblock {\em Phys. Rev. D}, 109(8):086015, 2024.

\bibitem{HotQCD:2019xnw}
H.~T. Ding et~al.
\newblock {Chiral Phase Transition Temperature in ( 2+1 )-Flavor QCD}.
\newblock {\em Phys. Rev. Lett.}, 123(6):062002, 2019.

\bibitem{Cai:2022omk}
Rong-Gen Cai, Song He, Li~Li, and Yuan-Xu Wang.
\newblock {Probing QCD critical point and induced gravitational wave by black
  hole physics}.
\newblock {\em Phys. Rev. D}, 106(12):L121902, 2022.

\bibitem{Bazavov:2020bjn}
A.~Bazavov et~al.
\newblock {Skewness, kurtosis, and the fifth and sixth order cumulants of net
  baryon-number distributions from lattice QCD confront high-statistics STAR
  data}.
\newblock {\em Phys. Rev. D}, 101(7):074502, 2020.

\bibitem{Borsanyi:2018grb}
Szabolcs Borsanyi, Zoltan Fodor, Jana~N. Guenther, Sandor~K. Katz, Kalman~K.
  Szabo, Attila Pasztor, Israel Portillo, and Claudia Ratti.
\newblock {Higher order fluctuations and correlations of conserved charges from
  lattice QCD}.
\newblock {\em JHEP}, 10:205, 2018.

\bibitem{Luo:2017faz}
Xiaofeng Luo and Nu~Xu.
\newblock {Search for the QCD Critical Point with Fluctuations of Conserved
  Quantities in Relativistic Heavy-Ion Collisions at RHIC : An Overview}.
\newblock {\em Nucl. Sci. Tech.}, 28(8):112, 2017.

\bibitem{Gao:2020qsj}
Fei Gao and Jan~M. Pawlowski.
\newblock {QCD phase structure from functional methods}.
\newblock {\em Phys. Rev. D}, 102(3):034027, 2020.

\bibitem{Shah:2024img}
Hitansh Shah, Mauricio Hippert, Jorge Noronha, Claudia Ratti, and Volodymyr
  Vovchenko.
\newblock {Locating the QCD critical point from first principles through
  contours of constant entropy density}.
\newblock 10 2024.

\bibitem{Hippert:2023bel}
Mauricio Hippert, Joaquin Grefa, T.~Andrew Manning, Jorge Noronha, Jacquelyn
  Noronha-Hostler, Israel Portillo~Vazquez, Claudia Ratti, Romulo Rougemont,
  and Michael Trujillo.
\newblock {Bayesian location of the QCD critical point from a holographic
  perspective}.
\newblock 9 2023.

\bibitem{Chen:2024epd}
Bing Chen, Xun Chen, Xiaohua Li, Zhou-Run Zhu, and Kai Zhou.
\newblock {Exploring Transport Properties of Quark-Gluon Plasma with a
  Machine-Learning assisted Holographic Approach}.
\newblock 4 2024.

\end{thebibliography}

\end{document}